\documentclass[journal]{IEEEtran}
\ifCLASSINFOpdf
\else
\fi

\usepackage{amssymb}
\usepackage{pifont}
\usepackage{makecell}
\usepackage{pdflscape}
\usepackage{booktabs}

\newcommand{\xmark}{\ding{55}}%
\newcommand{\cmark}{\ding{51}}%
\usepackage[dvipsnames]{xcolor}
\usepackage{epigraph}

\usepackage{xcolor}
\usepackage{mathtools}
\usepackage{amssymb}
\usepackage{amsmath}
\newcommand{\norm}[1]{\left\lVert#1\right\rVert}
\usepackage{todonotes}

\usepackage[acronym, shortcuts, nohypertypes={acronym}]{glossaries}
\usepackage[inline]{enumitem}
\newacronym{AE}{AE}{autoencoder}
\newacronym{AI}{AI}{artificial intelligence}
\newacronym{ASR}{ASR}{automatic speech recognition}
\newacronym{CNN}{CNN}{convolutional neural network}
\newacronym{CycleGAN}{CycleGAN}{cycle-consistent generative adversarial network}
\newacronym{DL}{DL}{deep learning}
\newacronym{DNN}{DNN}{deep neural network}
\newacronym{ESS}{ESS}{emotional speech synthesis}
\newacronym{ETTS}{ETTS}{emotional text-to-speech}
\newacronym{EVC}{EVC}{emotional voice conversion}
\newacronym{ExVo}{ExVo}{Expressive Vocalisations Workshop and Competition}
\newacronym{GAN}{GAN}{generative adversarial network}
\newacronym{GMM}{GMM}{Gaussian mixture model}
\newacronym{GRU}{GRU}{gated recurrent unit network}
\newacronym{GST}{GST}{global style token}
\newacronym{HCI}{HCI}{human-computer interaction}
\newacronym{HMM}{HMM}{hidden Markov model}
\newacronym{LSTM}{LSTM}{long short-term memory network}
\newacronym{ML}{ML}{machine learning}
\newacronym{MOS}{MOS}{mean opinion score}
\newacronym{NMF}{NMF}{non-negative matrix factorisation}
\newacronym{PPG}{PPG}{phonetic posteriorgram}
\newacronym{RNN}{RNN}{recurrent neural network}
\newacronym{S2S}{seq2seq}{sequence-to-sequence}
\newacronym{SER}{SER}{speech emotion recognition}
\newacronym{SPSS}{SPSS}{statistical parametric speech synthesis}
\newacronym{TTS}{TTS}{text-to-speech synthesis}
\newacronym{VAE}{VAE}{variational autoencoder}
\newacronym{VC}{VC}{voice conversion}

\usepackage[natbib, bibencoding=utf8, citestyle=numeric, bibstyle=ieee, maxbibnames=999, maxcitenames=2, mincitenames=1, sortcites]{biblatex}
\bibliography{bibtex/bib/dss.bib}

\usepackage[capitalise, nameinlink, noabbrev]{cleveref}

\newcommand{\eg}{e.\,g.}
\newcommand{\ie}{i.\,e.}

\begin{document}
%
\title{
An Overview of Affective Speech Synthesis and Conversion in the Deep Learning Era
} 

%
%
%

\author{
Andreas Triantafyllopoulos,~\IEEEmembership{Student Member, IEEE}, 
Bj\"orn W.\ Schuller,~\IEEEmembership{Fellow, IEEE}, 
G\"ok\c{c}e \.{I}ymen, 
Metin Sezgin,~\IEEEmembership{Member, IEEE}, 
Xiangheng He, 
Zijiang Yang,~\IEEEmembership{Student Member, IEEE}, 
Panagiotis Tzirakis~\IEEEmembership{Member, IEEE}, 
Shuo Liu, 
Silvan Mertes, 
Elisabeth Andr\'e,~\IEEEmembership{Senior Member, IEEE}, 
Ruibo Fu,~\IEEEmembership{Member, IEEE}, 
Jianhua Tao,~\IEEEmembership{Senior Member, IEEE}
\thanks{A. Triantafyllopoulos, B.\ Schuller, X.\ He, Z.\ Yang, S.\ Liu are with the Chair of Embedded Intelligence for Health Care and Wellbeing, University of Augsburg, Germany.}
\thanks{B.\ Schuller, P.\ Tzirakis are with GLAM -- the Group on Language, Audio, \& Music, Imperial College London, SW7 2AZ London, UK. }
\thanks{G.\ \.{I}ymen, Metin Sezgin are with the KUIS AI Lab., College of Engineering, Ko\c{c} University, Istanbul, Turkey. }
\thanks{S.\ Mertes, E.\ Andr\'e are the Chair of Human-Centered Artificial Intelligence, University of Augsburg, Germany. }
\thanks{R.\ Fu, J.\ Tao are with the Institute of Automation, Chinese Academy of Sciences, Beijing, China. }
\thanks{Manuscript received September 30, 2022}}

%
%

\markboth{Submitted to the Proceedings of IEEE}%
{Shell \MakeLowercase{\textit{et al.}}: Bare Demo of IEEEtran.cls for IEEE Journals}
%



\maketitle

\begin{abstract}
Speech is the fundamental mode of human communication, and its synthesis has long been a core priority in human-computer interaction research.
In recent years, machines have managed to master the art of generating speech that is understandable by humans.
But the linguistic content of an utterance encompasses only a part of its meaning.
Affect, or expressivity, has the capacity to turn speech into a medium capable of conveying intimate thoughts, feelings, and emotions -- aspects that are essential for engaging and naturalistic interpersonal communication.
While the goal of imparting expressivity to synthesised utterances has so far remained elusive, following recent advances in text-to-speech synthesis, a paradigm shift is well under way in the fields of affective speech synthesis and conversion as well.
Deep learning, as the technology which underlies most of the recent advances in artificial intelligence, is spearheading these efforts.
In the present overview, we outline ongoing trends and summarise state-of-the-art approaches in an attempt to provide a comprehensive overview of this exciting field.
\end{abstract}

\begin{IEEEkeywords}
Affective Computing, Speech Synthesis, Emotional Voice Conversion, Deep Learning.
\end{IEEEkeywords}

%
\IEEEpeerreviewmaketitle

\section{Introduction}
%
%
%
%

\epigraph{
We all have the capacity to be creative.
We're all driven to share our deepest dreams and ideas with the world.
When we think of the most talented creative people, they speak to us in a unique way.
A phrase we often hear is \emph{``Having a creative voice.''}
}{--- Val Kilmer}
\IEEEPARstart{T}{he} story of Val Kilmer, a world-renowned actor who lost his voice to throat cancer at the peak of his career, is a poignant reminder to the importance of verbal communication in human societies\footnote{A video of the reconstruction of Val Kilmer's voice for the purposes of Top Gun 2 by SONANTIC can be found in https://www.youtube.com/watch?v=OSMue60Gg6s.}.
A voice is more than the sum of its words; it is a conduit of one's individuality, their emotions, their unique worldview.
People who suffer from similar conditions understand that the mere verbalisation of their words using assistive technologies is not enough to give them back their voice.
They need to regain their lost \emph{emotional expressivity}~\citep{Fiannaca18-VAU}.

If artificial beings are ever able to attain an equal standing in human societies, why should they need any less?
While contemporary \ac{AI} research has set its sights to more attainable, down-to-earth goals, the long-standing dream of \ac{AI} researchers is to simulate, or perhaps overcome, human intelligence.
This goal may well require machines to have emotions as, to quote one of the forefathers of the field, Marvin Minsky~\citep{Minsky88-SOM}: ``The question is not whether intelligent machines can have any emotions, but whether machines can be intelligent without any emotions.''
And any being that has emotions requires an avenue to express them.

Affective computing is the subfield of \ac{AI} that concerns itself with the computational modelling, understanding, and expression of emotions~\citep{Picard00-AC}.
One of its primary goals is to facilitate more natural \ac{HCI} through the modelling of affect, which is a key component of human behaviour.
To that end, language, and, in particular, \emph{spoken} language, is the most natural form of communication.
If machines are ever to become natural conversational partners, they have to master the art of speech generation -- including the prosodic intonations attributable to the expression of affect~\citep{Batliner05-PMA}.
This is the domain of \emph{affective speech synthesis}, a computational paradigm which attempts to generate realistic-sounding affective speech.
We define affective speech synthesis as a sub-field of voice transformation~\citep{Stylianou09-VT}, which corresponds to the modification of all potential parameters of speech, and as a super-field of \ac{ESS}, which corresponds to the modification of emotion.
Affective speech synthesis, in contrast, extends beyond emotions by covering all aspects that fall under the umbrella of \emph{computational paralinguistics}~\citep{Schuller13-CP}, such as mood, personality, and social status.
Nevertheless, as emotion remains the primary pursuit of current research efforts, we will concentrate our review on this aspect of affective synthesis.
We also consider both the general case of synthesising an affective utterance directly from the input text, as well as that of modifying a neutral one to capture the desired emotion -- a subfield of synthesis generally referred to as affective, or emotional, voice conversion.

The first attempts to infuse emotion into synthesised speech were made before the field's name was even coined~\citep{Murray89-HAMLET, Cahn89-AE, Cahn90-AE,Kitahara92-PCT,Granstrom92-TUO,Murray93-TTS}.
For a long period, research on emotional speech synthesis and conversion have primarily focused on rule-based approaches guided by experts and listening experiments\footnote{An online `museum' including listening examples of most such attempts is found at: http://emosamples.syntheticspeech.de/}.
This is in contrast to \ac{SER} --the `opposite end' of synthesis-- which has been dominated by a data-driven paradigm~\citep{Schuller18-SER}.

The last few years have seen tremendous progress in the `sister fields' of speech synthesis and voice conversion.
The landmark work of \citet{Oord16-WAG} revolutionised the field of \ac{TTS}, signalling the advent of the \ac{DL} era and, more generally, solidifying the switch to a data-driven paradigm, where a mapping from text to speech is \emph{learnt} using data.
Similar approaches are now spearheading research in affective speech synthesis as well~\citep{Schuller18-SER}.

\ac{TTS} approaches have reached such performance levels that the task is considered by many to be `solved' -- layman users in particular expect commercial \ac{TTS} systems to work flawlessly, as seen, for example, in the recent wave of voice assistants.
Accordingly, \ac{TTS} research has exploded in recent years.
In contrast, the field of affective speech synthesis has attracted somewhat less attention in the field of \ac{HCI}, which is nevertheless substantially increasing.
Yet, even though significant progress has been made in that area as well, the goal of human-level, controllable emotional expressivity still remains elusive.

In an attempt to summarise recent efforts, synthesise existing approaches, identify missing gaps, and highlight promising research directions, we have construed a literature review of \emph{deep}, \emph{affective} speech synthesis and conversion methods. 
Our overview thus fills the gap between recent surveys in deep speech synthesis
which focus on `mere' \ac{TTS}~\citep{Tan21-NSS}, and older affective speech synthesis reviews which have become 
largely 
obsolete in the deep learning era~\citep{Schroeder01-ESS}, or newer ones which are more limited in scope~\citep{Zhou22-ESD, Yang22-S2S}.

The remainder of this work is structured as follows: We first present an overview of where affective speech synthesis fits in an affective computing application.
We then give a brief introduction on deep speech synthesis in general, followed by a thorough review of deep emotional speech synthesis and conversion.
Finally, we summarise major observations and outline potential avenues for future research.

\section{Affective speech generation}

In this section, we begin by giving a definition of what is entailed by affective speech generation.
We use the term ``generation'' to encompass all aspects of a process that begins with a `decision' on what emotion needs to be generated, a selection of the appropriate text, and the final \emph{synthesis} of the waveform as the last step.
As affect is an overloaded term, we first give a concrete definition of it for the purposes of our review.
After defining what we mean by it, we continue our overview with a short introduction on the implicit model which underlies all affective computing applications: that of an agent who is able to respond emotionally to external stimuli.
This agent may be fully artificial, as in the case of an autonomous conversational agent that interacts with humans, or `hybrid' in the case of an emotional voice conversion system that augments the capabilities of speech-impaired individuals~\citep{Fiannaca18-VAU}.
We then introduce the module of that agent responsible for the generation of affect in speech, followed by an introduction of the different representation models used in the computational modelling of affect.
Finally, we introduce the acoustic correlates of affect which have guided the related speech research for decades.

\subsection{What is affective speech?}
We begin with a definition of the concepts used here as, due to their subjective nature, they are often conflated with one another.
We use the term ``affect'' in its broadest connotation, as it was introduced in the inaugural work of Picard~\citep{Picard00-AC}.
Specifically, we go by the definition of affect ``as a broader term, encompassing all kinds of manifestations of personality such as mood, interpersonal stances, or attitudes''~\citep{Schuller13-CP}.
This differs from standard psychological interpretations which more narrowly define affect as the manifestation of a subjectively experienced emotion~\citep{Munezero14-AFE}.
Thus, in our review, affective speech is speech which encapsulates all possible paralinguistic traits and states~\citep{Schuller13-CP}.

Seen from that perspective, the majority of research in affective speech synthesis has been actually devoted to \ac{ESS}, with considerably less emphasis on personality and other states or traits~\citep{Schroder11-BAS, Schuller13-CP}.
This discrepancy is even more pronounced in the ongoing \acl{DL} era, with far more work devoted to \ac{ESS} than any other aspects of affect.
For this reason, we focus our review on \acl{ESS} and leave a discussion of other aspects of affect for \cref{ssec:future}.

We note that several recent works investigate `expressive' \ac{TTS}~\citep{Huybrechts21-LRE, Lakhotia21-GSL, Polyak21-SRD, Kharitonov22-TFPA, Liu22-ETTS}.
This term is increasingly appearing on \ac{TTS} papers to mean the control of prosody and other factors not related to content.
However, it is often the case that this control is applied manually to change the style of the synthesised utterance.
It is thus missing the explicit link to emotion that we consider critical for \ac{ESS}.
Naturally, prosody is highly related to emotion; therefore, any manipulation of it might result in a change in the perceived emotion.
Yet, expressive \ac{TTS} methods tend to leave out any evaluation of the emotional content of the synthesised utterances.
For this reason, we only tangentially refer to them in our review.
We also note that in the earliest affective speech synthesis papers, expressivity was considered a synonym to affect~\citep{Cahn89-AE}.
Indeed, the mechanisms used to induce expressivity are almost identical to those used to induce emotionality, albeit the latter explicitly use emotional information for guidance.
We consider this a necessary prerequisite for successful \ac{ESS}.

\subsection{Affective agent model}
\begin{figure*}
    \centering
    \includegraphics[width=\textwidth]{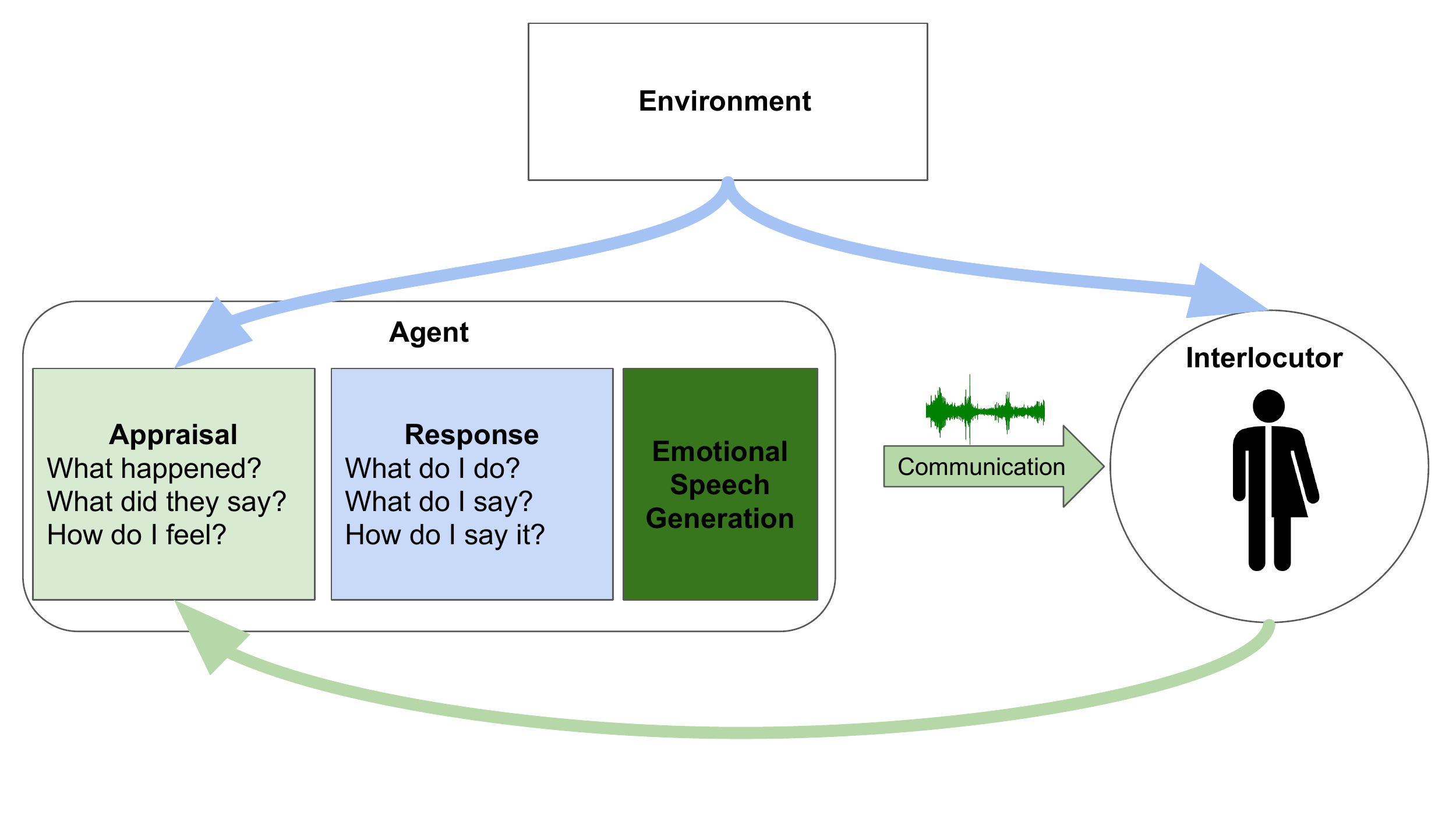}
    \caption{
    Overview of our emotional speech generation model.
    We assume the presence of an artificial agent who can receive inputs from the environment (including responses from their interlocutor) and proceed to appraise the situation (an appraisal can be either hardcoded or learnt) and generate an appropriate response.
    This response is then converted to an appropriate speech signal by an emotional speech generation module, which is the core focus of this review, and transmitted to the interlocutor.
    }
    \label{fig:generation}
\end{figure*}

While the majority of affective speech generation works are concerned with the task of endowing a synthesised utterance with appropriate emotional inflections, this is but the last step in the pipeline of an affective agent\footnote{With the word `agent' here we mean a software component which simulates a desired behaviour in any digital application, not necessarily an embodied conversational agent.}.
\cref{fig:generation} shows a coarse model of what is at play in an affective computing application.
The agent, rather than existing in a vacuum, is embedded in an environment (\eg, its application or, even, the entire world) and interacts with an interlocutor (usually a human; in the future, potentially other artificial entities)~\citep{Schroder11-BAS}.
It receives inputs from this environment --including responses/queries by its interlocutor-- and generates an appropriate response.

An important step in this process is the appraisal of all input stimuli.
According to appraisal theory~\citep{scherer1999appraisal, ellsworth2003appraisal}, inputs from the environment are evaluated with respect to the agent's goals and concerns along several dimensions.
For example, \citet{ellsworth2003appraisal} proposed novelty (how much new information was contained in the stimulus), intrinsic pleasantness/valence (how positive or negative the stimulus `feels' for the agent), relevance (how pertinent is the information to its goals), urgency (how fast it needs to respond), and power/control (how much is the situation under its control).
While this list is not exhaustive, and alternative appraisal theories have been proposed over the years, we believe it captures a core component of an affective agent: namely, that the appropriate text and emotion to be synthesised have to be somehow defined.

Of course, in most contemporary affective computing applications, the appropriate response is dictated by the creator of the application.
Most of them contain hardcoded behaviours that the agents must follow (\eg, be constantly `happy' or `pleasing').
However, some recent works are already experimenting with learnt behaviours -- for example, using reinforcement learning to train a dialogue agent to incorporate emotional responses~\citep{Lan21-EDR}, as these have been shown to increase subjective scores of dialogue richness~\citep{Chiba18-NTO, Lubis18-EPE, Bucinca20-AffectON}.
As the field progresses, we expect more research towards less hardcoded and more learnt (or emerging) behaviours.

\subsection{Emotional speech generation}

\begin{figure*}
    \centering
    \includegraphics[width=\textwidth]{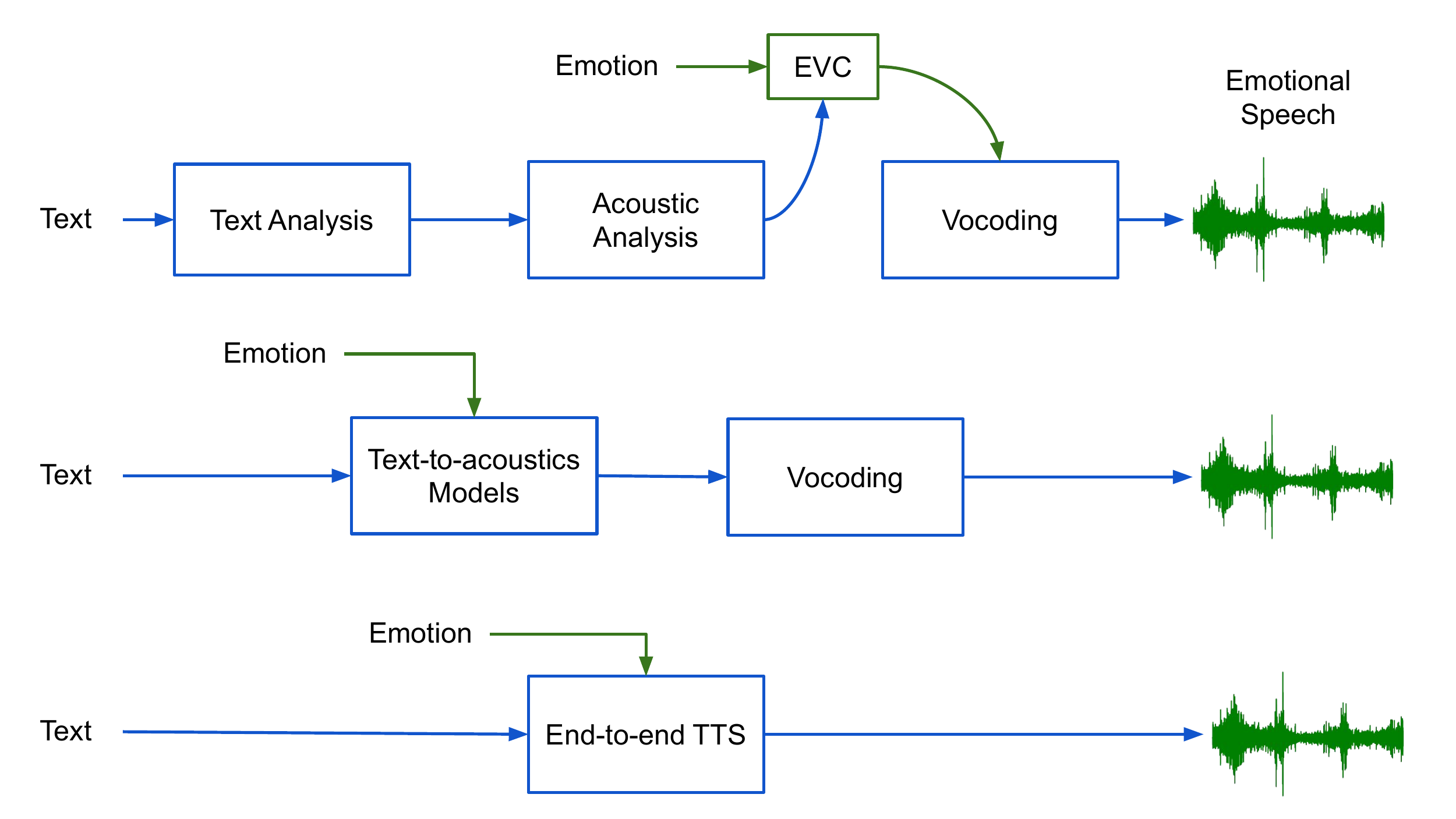}
    \caption{
    Overview of an emotional speech synthesis module.
    Emotional synthesis (green) is superimposed on \ac{TTS} pipelines (blue), which traditionally consist of 3 steps (top): text analysis, acoustic analysis, vocoding (synthesis).
    In the standard setup, emotion is used to modulate the acoustic features before vocoding (\acf{EVC}).
    Deep learning models typically incorporate two (middle) or even all (bottom) of these steps in a single model.
    In this case, emotion is used as extra, conditioning information to inform the generation of the respective outputs of each model.
    }
    \label{fig:ETTS}
\end{figure*}
For the present, we focus on the last step of an affective speech generation agent, which is the synthesis of the speech utterance itself, after a suitable text and emotion response have been determined by other processes~\citep{Bucinca20-AffectON}.
An overview of this process is shown in \cref{fig:ETTS}, where we present the common blocks of a \ac{TTS} system and the adaptations required to enrich it with emotion.
In brief, a \ac{TTS} system (blue boxes and lines) comprises three steps:
\begin{enumerate*}
    \item a text analysis module that converts the input text to appropriate linguistic features,
    \item an acoustic model that converts those features to acoustic features, and,
    \item a vocoder, which generates the final utterance.
\end{enumerate*}
While this is the traditional approach to \ac{TTS}, the barriers between the different steps have began to erode with the advent of \ac{DL}, with a single architecture often subsuming several (or even all) steps.
We review both traditional, multi-step synthesis and \ac{DL}-based synthesis in \cref{sec:tts}.

Incorporating emotion to this pipeline is primarily done in two ways (green boxes and lines): Either an \ac{EVC} module is tasked with adapting the emotion of the synthesised speech, or the transformation is made as an intermediate step before vocoding.
Due to the recent success of \ac{TTS} architectures, most \ac{ESS} works are actually performing \ac{EVC}; however, there are several works which go directly from phoneme sequences to acoustic features, thus incorporating the first two steps of a synthesis pipeline.
All these methods will be reviewed in \cref{sec:ess}.
Naturally, the target emotion may influence the generation of the text response itself, but as previously mentioned, we ignore this step for our purposes.

\subsection{Computational models of affect}
As \citet{fehr1984concept} famously wrote: ``Everyone knows what an emotion is, until asked to give a definition. Then, it seems, no one knows.''
However, to generate an emotion, one must nevertheless have a proper representation of it.
Several different emotion theories have emerged over the years, each focusing on different, but oftentimes related, aspects of emotion~\citep{scherer2019towards}.
Two of those have dominated the computational modelling of emotion~\citep{Schuller18-SER}: discrete emotion theories, where emotions are considered to fall under discrete categories like Ekman's big six~\citep{ekman1992argument}, and dimensional ones, like Russel's arousal, valence, and dominance~\citep{russell1977evidence}.
Most affective speech synthesis works have adopted the first formulation, and assume emotion to come in discrete categories, thus transforming one to the other (or neutral to one of them), while only a few pursue the synthesis of dimensional affect instead~\citep{Schroder06-DoA, xue2018voice}.
Our review will accordingly focus on categorical \ac{ESS} methods, as these have dominated the ongoing \ac{DL} era.


\subsection{Acoustic correlates of affect}
\label{ssec:correlates}
Affective speech synthesis is concerned with adapting those constituents of a speech signal that convey affective information.
Thus, progress in this field depends on progress in the mirror field of affect recognition and analysis, where considerably more research has been invested in the last decades~\citep{Banse96-APV, Johnstone00-VCE, Schuller18-SER}.
Speech parameters that are identified as being conducive to the recognition of affect in speech are readily co-opted by researchers to control affect during synthesis -- and vice versa~\citep{Batliner05-PMA}.

A large body of literature has linked the manifestation of affect in speech to suprasegmental features like prosody, voice quality (\eg, jitter and shimmer), spectral and energy features, and temporal patterns such as tempo and pausing~\citep{Banse96-APV, Johnstone00-VCE}.
For example, anger was shown to correspond to a higher mean F0 and energy, while ``hot'' anger also induced a higher variability and range of F0~\citep{Banse96-APV}.
These features in turn became the main knobs twisted by early affective speech synthesis models to achieve their required results~\citep{Cahn89-AE, Cahn90-AE, Murray89-HAMLET}.
Recent approaches have attempted to substitute them with learnt representations, in the hope that those are better able to capture emotional information~\citep{Choi21-S2S, Zhou22-EIC}.
Nevertheless, those features have left their mark on affective speech synthesis research as several works still use them --in some form-- to guide the generation of emotional utterances~\citep{Lorenzo-Trueba18-IDR, Shankar19-MSE, Luo19-EVC, Schnell21-EII, Zhang22-iEmoTTS}.
These approaches therefore constitute the main focus of our review.

There has also been some interest in other aspects of vocalisations impacted by affect.
For example, \citet{tahon2018can} investigated the potential of generating emotional pronunciations to improve expressivity.
More recently, \citet{Baird22-EXVO} launched the \ac{ExVo} to foster more research in the generation of realistic emotional ``vocal bursts''~\citep{Cowen19-BHV}.
Combining such approaches with the synthesis methods removed here has great potential to improve the expressivity and emotionality of generated utterances, and we will discuss the potential of such attempts in \cref{ssec:future}. 

\section{Speech synthesis}
\label{sec:tts}

The goal of a speech synthesis system, also known as \acl{TTS}, is to generate artificial, human-like speech from a given text input.
Speech synthesis is, thus, naturally the backbone of affective speech synthesis, as the generation of realistic-sounding utterance is a prerequisite for enhancing its expressivity.
The first recorded \ac{TTS} system is Wolfgang von Kempelen's 18th century pipes and bellows machine, which was able to produce vowel and consonant approximations, that, when properly combined, allowed visitors in his lab to recognise certain words~\citep{Dudley50-WvK}.
The field has obviously progressed a lot from those early origins with the introduction of digital technology.
Earlier digital attempts at \ac{TTS} include articulatory, formant, and concatenative synthesis.
The field then inherited advances in statistical machine learning and transitioned to the \ac{SPSS} paradigm, whose influence is still ripe throughout contemporary DL-based \ac{TTS} systems.

As the development of \ac{ESS} has developed in tandem with that of \ac{TTS}, we considered a brief overview of early synthesis methods necessary.
This is followed by a review of \emph{deep} speech synthesis methods in \cref{subsec:deep-tts}, which sets the tone for our deep affective speech synthesis overview presented in the next section.
This section is concluded with an overview of (deep) voice conversion, an application field of voice transformation which has a lot in common with \ac{ESS}~\citep{Stylianou09-VT}.

\subsection{A brief history of speech synthesis}
\label{subsec:early-tts}

The earliest (digital) \ac{TTS} systems attempted to simulate the human articulatory system by creating models for the movement of lips, tongue, glottis, and vocal tract -- thus not differing much in spirit from the mechanical apparatus of von Kempelen.
This came to be known as \emph{articulatory synthesis}~\citep{Coker76-MAD}.
This paradigm was met with severe challenges in the modelling of articulatory behaviour, and was abandoned for a simpler, source-filter model that lends itself better to parameter control: \emph{formant synthesis}~\citep{Allen79-MIT, Klatt87-TTS}.
This type of synthesis relies on a rule-based modification of the formant amplitudes and frequencies of an excitation signal to produce the required utterance.
These rules are derived by linguistic analysis.
While this system has more modularity than articulatory synthesis, the difficulty in identifying an appropriate set of rules has led to its abandonment in favour of data-driven paradigms.

To overcome the challenges associated with building a proper articulatory model or assembling a complete list of formant rules, the community next turned to \emph{concatenative speech synthesis}, where the target utterance is constructed from a set of pre-recorded building blocks: words, syllables, half-syllables, phonemes, diphones, or triphones~\citep{Moulines90-PSW}.
These pre-recorded units are concatenated to produce the utterance of interest.
However, concatenative synthesis suffers from discontinuity effects (as the prosody of each recording can differ), and a dramatic increase in the data needed to cover all combinations of units.

All these downsides led to the adoption of a learning paradigm in the name of \emph{\acf{SPSS}}~\citep{Tokuda00-SPG, Zen09-SPSS}.
\ac{SPSS} adopts the three-stage model presented in \cref{fig:ETTS}, namely the use of text analysis to suitable linguistic representations of the target utterance, the prediction of speech parameters using an acoustic model, and the final waveform synthesis (vocoding).
In particular, the \emph{text analysis} module includes necessary pre-processing steps (text normalisation, grapheme-to-phoneme conversion, etc.) followed by the extraction of all relevant features, like phonemes, duration, or part-of-speech tags.
Those features, along with the accompanying speech parameters, are fed to a statistical \ac{ML} model that learns a mapping from linguistic to acoustic features (\eg, the fundamental frequency, spectrum, or cepstrum).
Due to the sequential nature of this data, \acp{HMM} have excelled at this type of modelling~\citep{Tokuda00-SPG}.
Finally, the acoustic features are propagated to a suitable vocoder for the synthesis step.
Some notable vocoders are WORLD~\citep{Morise16-WORLD} and STRAIGHT~\citep{Kawahara06-STRAIGHT}.
It is important to emphasise that several (even all) of those steps are \emph{learnable from data} -- which is precisely what gave this family of methods its name.
Specifically, to learn any of the mappings from graphemes to phonemes to acoustics to waveform, matching data (\ie, matching text and speech pairs, often obtained from several speakers and of high amount) is needed.
This fundamental attribute of \ac{SPSS} is what makes it the forefather of modern-day deep speech synthesis methods.

\subsection{Deep speech synthesis}
\label{subsec:deep-tts}

\begin{figure*}
    \centering
    \begin{tabular}{c|c}
    \includegraphics[width=.49\textwidth]{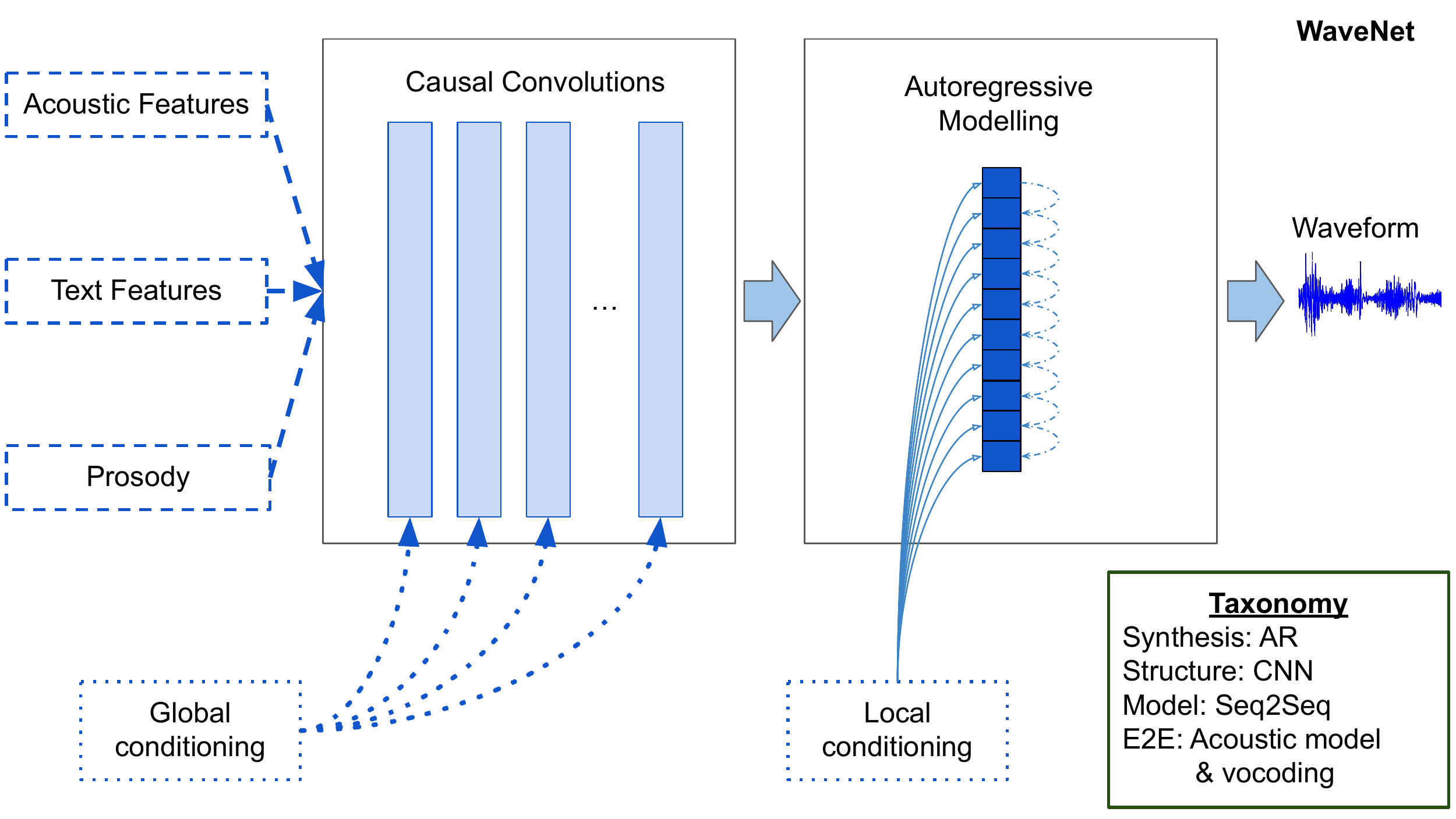}&
    \includegraphics[width=.49\textwidth]{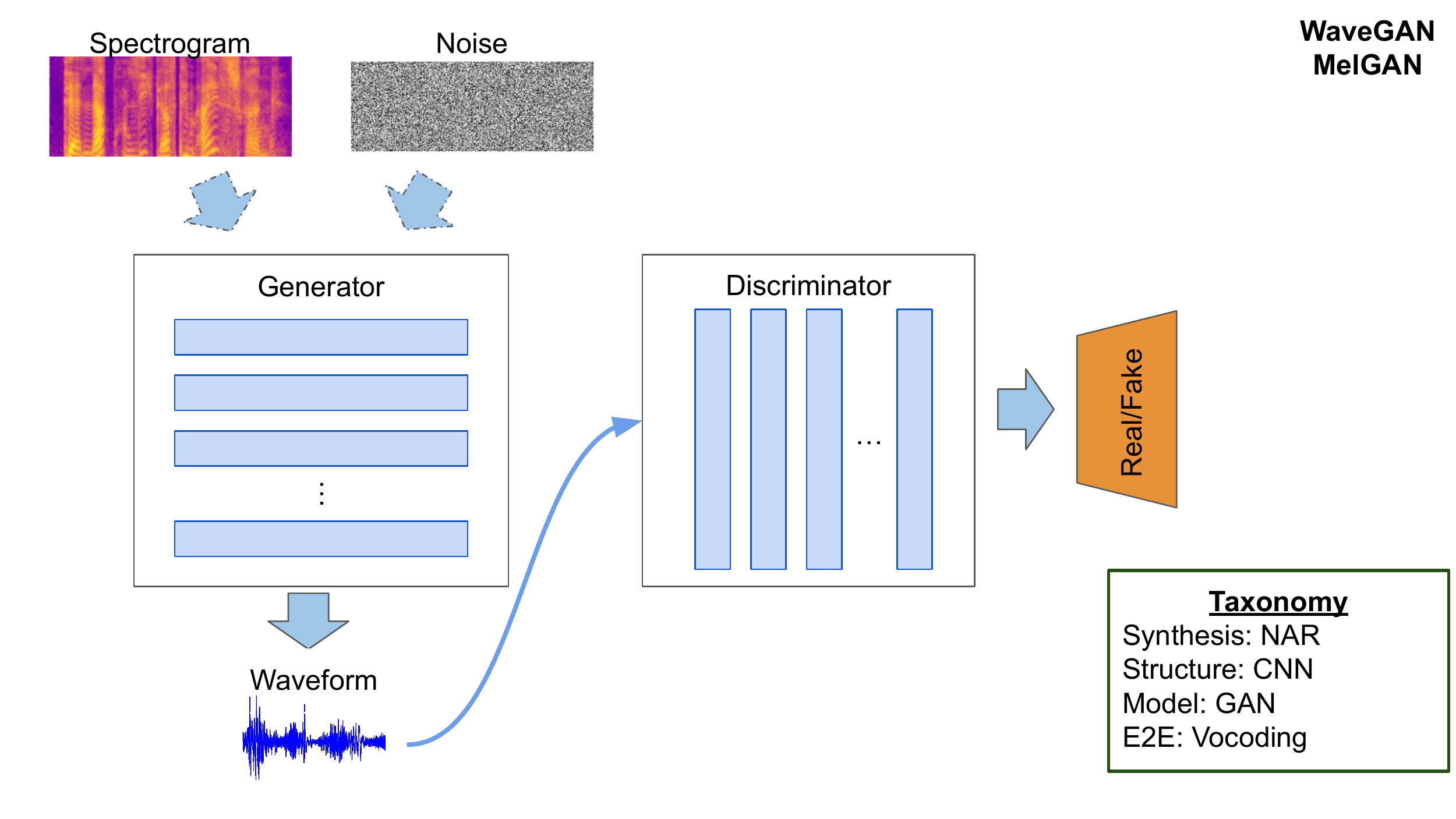}\\
    \hline
    \includegraphics[width=.49\textwidth]{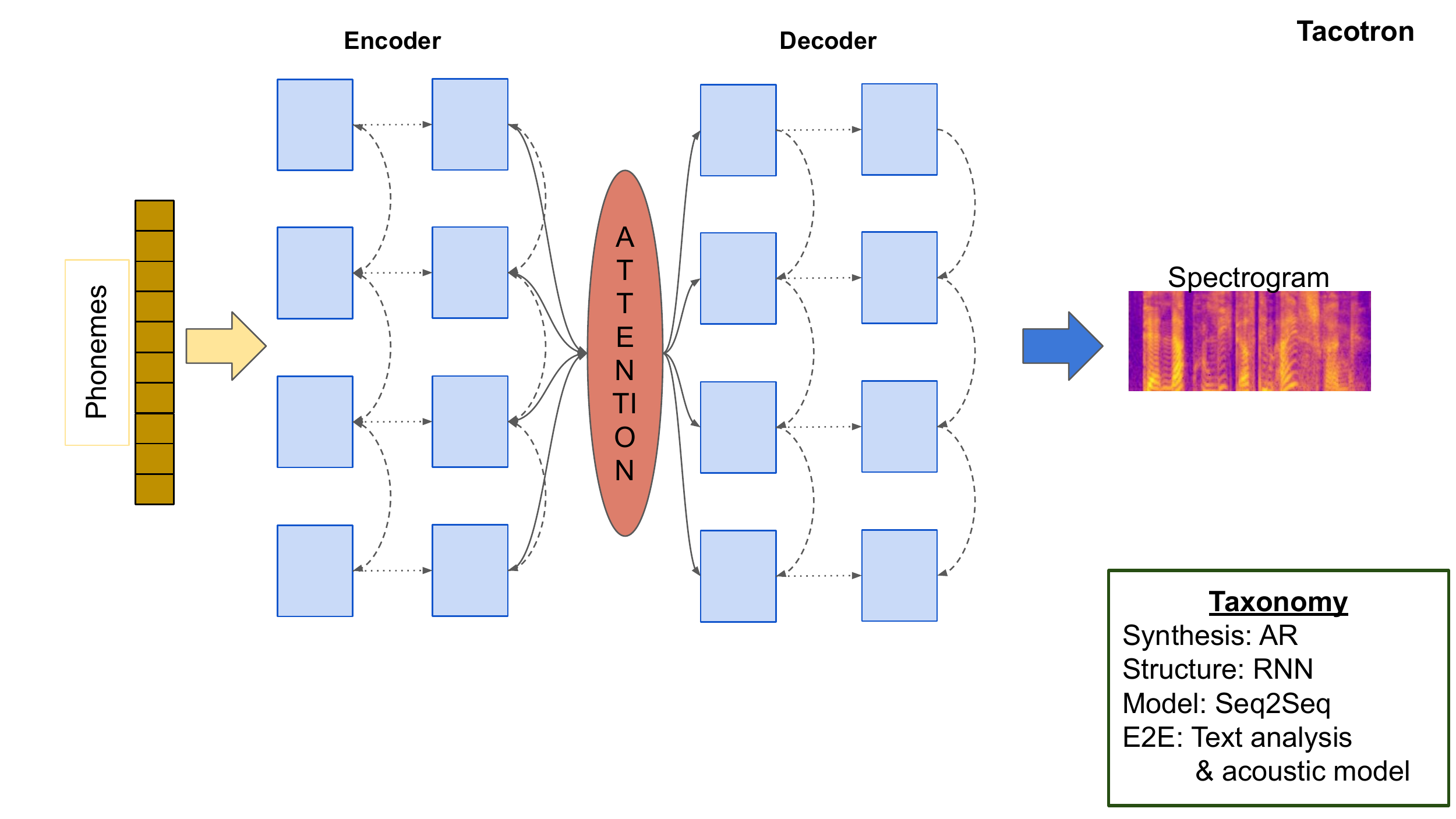}&
    \includegraphics[width=.49\textwidth]{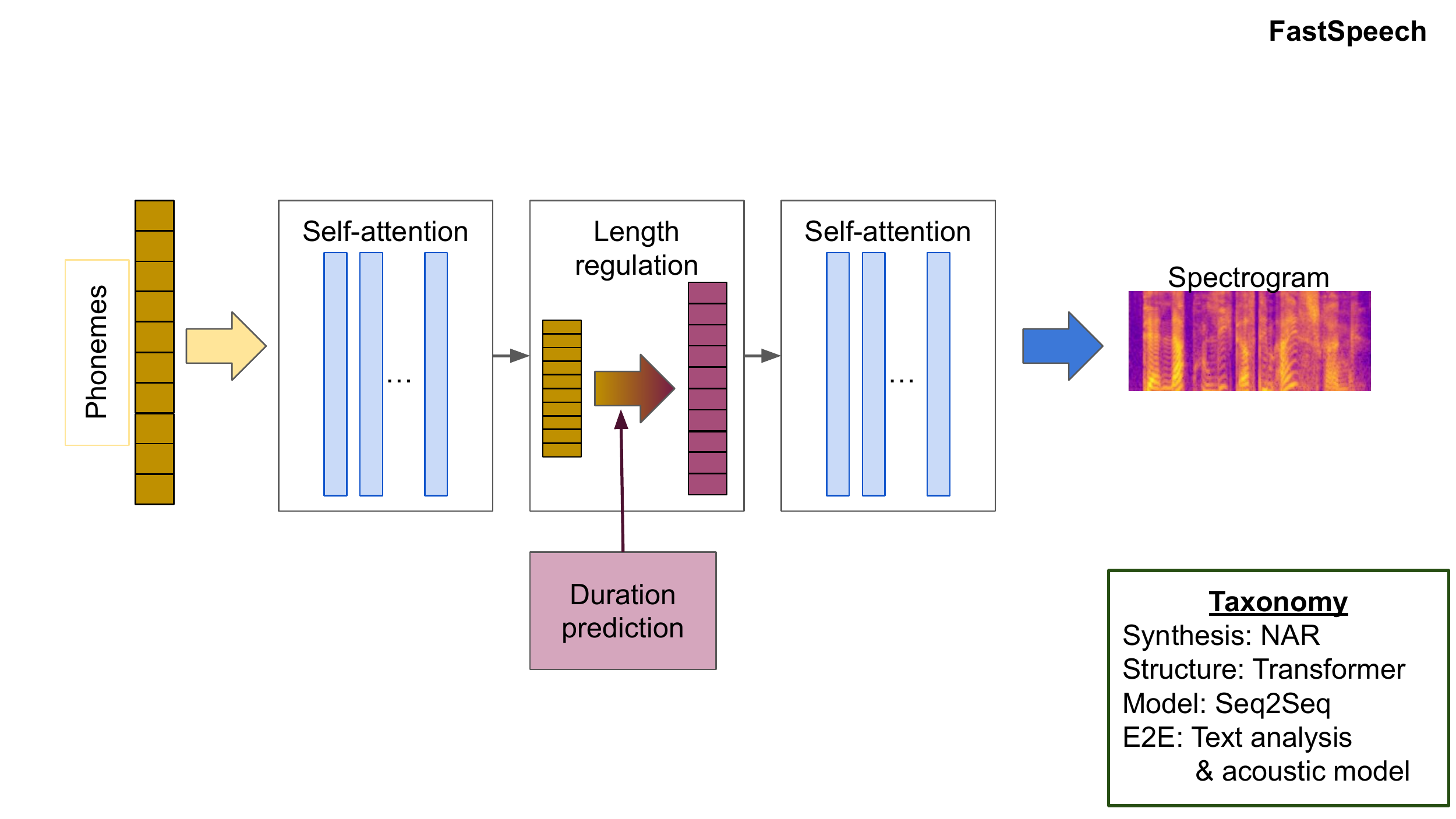}\\
    \end{tabular}
    \caption{
    Overview of main deep \ac{TTS} paradigms.
    WaveNet was first introduced as a text-to-waveform model (thus combining an acoustic model with vocoding), which could be locally and globally conditioned on additional information; it was later extended to synthesise waveforms from input spectrograms, thus relegated to the role of a traditional vocoder.
    \acp{GAN} are commonly used to map spectrograms to waveforms (effectively acting as vocoders), or to `imagine' waveforms from a random input, as such subsuming all intermediate steps of a \ac{TTS} pipeline as well as the mechanism to decide what text to output.
    Tacotron utilises \ac{S2S} models to learn a mapping from phonemes/characters to audio features, thus implicitly combining text analysis with an acoustic model; FastSpeech iterated on that by substituting \acp{RNN} with Transformers.
    }
    \label{fig:TTS}
\end{figure*}

Deep neural network-based synthesis co-opts neural networks as the models of choice to substitute one or more components of a traditional \ac{SPSS} pipeline.
First attempts usually centred around substituting \acp{HMM} with sequential models (\acp{RNN}~\citep{Jordan97-SOP} or \acp{LSTM}~\citep{Hochreiter97-LSTM}) for acoustic modelling, such as the early DeepVoice systems~\citep{Arik17-DVR, Gibiansky17-DV2}.
WaveNet was the first neural model to directly generate the waveform from linguistic features~\citep{Oord16-WAG}.
This was later followed by models trying to go directly from character/phoneme sequences to audio, like Tacotron~\citep{Wang17-TTE, Shen18-NTS}.
Nevertheless, several \ac{DL}-based methods are still using the traditional 3-step pipeline, but substitute intermediate steps with their \ac{DL} counterparts.
The defining characteristics of deep speech synthesis are thus threefold: a) methods follow the \ac{SPSS} formulation, b) all methods utilise deep neural networks in some, or all, steps of their pipeline, and c) some methods subsume some or all of the intermediate step in a single model -- these are the so-called end-to-end approaches.

\ac{DL}-based methods can be taxonomised along several categories:
\begin{itemize}
    \item \textbf{Autoregressive} (AR)~\citep{Oord16-WAG} vs \textbf{non-autoregressive} (NAR) structures~\citep{Peng20-NAN}.
    \item Type of \textbf{network structure}, where we primarily differentiated between \acp{CNN}~\citep{Arik17-DVR, Gibiansky17-DV2}, sequential models (\acp{RNN}, \acp{GRU}, \acp{LSTM})~\citep{Wang17-TTE, Shen18-NTS, Shen21-Tacotron2} which may or may not include attention, and self-attention models (\ie, Transformers)~\citep{Li18-NSS, Ren19-Fastspeech, Ren20-FS2}.
    \item Type of \textbf{generative model} (\eg, \ac{VAE}~\citep{Kingma13-VAE}, \ac{GAN}~\citep{Goodfellow14-GAN}).
    \item \textbf{Degree of end-to-end} behaviour, which is characterised by the steps of the traditional \ac{SPSS} pipeline that one or more (jointly trained) models subsume.
\end{itemize}
While such a categorisation is useful for differentiating between different \ac{TTS} approaches --and later on understanding \ac{ESS} ones-- it is important to stress that the boundaries between those categories are fluid and constantly changing.
For example, while WaveNet was first introduced as an autoregressive model which generates a waveform directly from linguistic features~\citep{Oord16-WAG}, thus integrating the acoustic model and vocoding aspects of an \ac{SPSS} pipeline, it was later extended to non-autoregressive synthesis~\citep{Oord18-PWF} and changed to produce speech conditioned on (Mel-)spectrograms rather than linguistic features~\citep{Shen21-Tacotron2}.
These rapid changes are expected as researchers continually optimise their pipelines in their quest for end-to-end synthesis.
Nevertheless, as our focus is on presenting the core ideas that have revolutionised the \ac{TTS} field in the last decade, we will primarily categorise approaches based on their earliest iterations.

An overview of recent, key \ac{TTS} contributions from the deep learning era is shown in \cref{fig:TTS}.
As previously mentioned, WaveNet~\citep{Oord16-WAG} was the first neural model to be proposed for speech synthesis.
In its first introduction, it was conceptualised as a mapping from textual and prosodic features to a raw waveform -- thus integrating the last two steps of an \ac{SPSS} pipeline.
WaveNet also introduced two key innovations in the field of audio modelling: a) The use of dilated convolutions, which allowed it to increase its receptive field and model long-range interactions. b) The ability to globally and locally condition the generation process, which proved instrumental in controllable \ac{TTS}, as well as emotional \ac{TTS} and voice conversion.
Follow-up iterations adapted the model to accept (Mel) spectrograms as input~\citep{Shen21-Tacotron2}, thus effectively transforming it into a more traditional vocoder.

Tacotron~\citep{Wang17-TTE} approached neural \ac{TTS} by combining the two frontends of the \ac{SPSS} pipeline, text analysis and acoustic modelling, using an encoder-attention-decoder framework.
By relying on \ac{S2S} models, optionally augmented with attention, Tacotron learns a mapping from phonemes/characters to spectrograms.
These spectrograms are then fed into a suitable vocoder; for that purpose, Tacotron1 used Griffin-Lim~\citep{Griffin84-SEM} whereas Tacotron2 used WaveNet~\citep{Shen21-Tacotron2}.
Due to the sequential nature of the encoder and decoder, the Tacotron series suffers from slower processing times and difficulties in addressing long-range dependencies.

Following the recent successes of self-attention architectures in modelling such dependencies~\citep{Vaswani17-AYN}, and their ability to generate their output in non-autoregressive fashion by processing their inputs in parallel, Transformers were introduced as an alternative to \acp{RNN} in the FastSpeech series~\citep{Ren19-Fastspeech, Ren20-FS2}.
FastSpeech relies on a series of Transformer blocks for encoding the input text sequence; another series of blocks decodes it to the output acoustic features that then serve as input to a suitable vocoder.
While Transformers have the advantage of processing the entire sequence in parallel, thus reducing runtime during inference, they require some adaptations to handle the problem of mismatched sequence lengths, as target acoustic features typically have a much longer duration than the input text.
This is handled by a duration prediction network, which is trained to predict this mismatch and upsample the learnt representations of the encoder to the necessary length before propagating them to the decoder.
FastSpeech2 is also trained to jointly predict the pitch and energy of the target speech, which is then used to further modulate the learnt representations of the encoder during inference and improve expressivity; this `variance adaptation' mechanism can be readily co-opted for \ac{ESS} by using it to inject emotional information as well.

Finally, another key contribution to the zoo of neural \ac{TTS} approaches is the introduction of \acfp{GAN}.
Following the seminal work of \citet{Goodfellow14-GAN}, \acp{GAN} have become mainstays in image, video, and audio generation.
For \ac{TTS}, there are two main categories of \acp{GAN}.
The first one is \ac{GAN} vocoders, whose generators accept as input spectrograms and output the raw waveform, with the waveform subsequently probed via the discriminator for its `realness'.
Key examples of this category are MelGAN~\citep{Kumar19-MGA}, Parallel WaveGAN~\citep{Yamamoto20-PWA}, HiFi-GAN~\citep{Kong20-HiFi} and others.

Sticking closer to the original formulation by \citet{Goodfellow14-GAN}, the second category includes models like WaveGAN~\citep{Donahue18-AAS}, which attempt to generate realistic speech from random inputs.
They thus effectively substitute the entire speech generation model --including the selection of the appropriate text to output-- with a single model.
While such methods are certainly intriguing, their opaqueness and lack of controllability make them unsuitable for current \ac{TTS} needs; still, it is an interesting avenue to explore in the search of models that can decide for themselves what they want to say.

We end this section with a note that we have omitted several key advances in deep speech synthesis.
As our goal was not to provide a comprehensive overview, but instead a brief one of core novelties introduced in the deep learning era, we have focused on those most pertinent to emotional synthesis.
Thus, among others, we have excluded flow-based~\citep{Prenger19-FBG} and diffusion-based models~\citep{Popov21-Grad}.
For a thorough review of neural speech synthesis which also includes these advances, we refer the reader to \citet{Tan21-NSS}.

\subsection{Deep voice conversion}


\Acf{VC} is the task of making a speech utterance from a \emph{source} speaker sound like it came from a \emph{target} speaker, while keeping the linguistic content unchanged. 
To further differentiate it from \ac{EVC}, we also require \ac{VC} to leave the emotion of the utterance unchanged.
Since \ac{VC} and \ac{EVC} share many commonalities, we defer a thorough consideration of speech conversion methods to \cref{sec:ess}.
Nevertheless, we provide a short synopsis of \ac{VC} methods here, as it is a vibrant sub-field of speech synthesis, with approaches first introduced there and later applied to \ac{EVC} and vice versa.
Two recent comprehensive reviews of \ac{VC} can be found in \citet{Mohammadi17-VC} and \citet{Sisman20-VC}.

Several attributes should be manipulated to make the speech of one individual sound like that of another.
The first one is the choice of words themselves.
Different people use different vocabularies and styles of speaking~\citep{Kinnunen10-TIS, Sisman20-VC}; therefore, to effectively transform the `identity' of a speech utterance, one should begin with the words that constitute it.
However, as we will later also ignore changes to vocabulary necessitated by changes in emotion, we also ignore this important aspect of voice conversion as well.
Instead, we focus on the other two attributes: Supra-segmental features like prosody and segmental ones like spectrum and formants.
Short-time spectral features are correlates of \emph{timbre}, which captures the `tone' of an utterance and is related to the physiological characteristics of the speaker~\citep{Kinnunen10-TIS}.
Prosody captures both physiological characteristics and speaking style~\citep{Kinnunen10-TIS}.
For this reason, several \ac{VC} works consider only timbre; this, however, limits the success of those methods as human impersonators are found to adapt their prosody as well~\citep{Kinnunen10-TIS}.

\ac{VC} research has a history of more than 30 years.
Early approaches utilised articulatory synthesis, but synthesised speech using the parameters of the target speaker~\citep{Childers89-VC}.
More recent attempts used \acp{GMM}~\cite{Stylianou98-CPT}, exemplar-based frameworks based on \ac{NMF}~\cite{Sisman18-AVC}, and \acp{HMM}~\cite{YamagishiONIK06}. 
However, despite several attempts, no notable progress was made until recent years, which saw advent of \acp{DNN}.
Recently proposed methods exploit the representation power of \acp{DNN} by means of \acp{VAE}~\citep{Saito18-NVC, Hsu16-VCF}, \acp{GAN}~\citep{Kameoka18-SNM, Kaneko18-CNV}, and \ac{S2S} models~\citep{LiuCKHL00M20, TanakaKKH19}.

A key distinction of \ac{VC} approaches is between those who use \emph{parallel} and \emph{non-parallel} training data.
While this distinction will be further elucidated in our discussion of \ac{EVC}, where it plays an equally crucial role, we already need to touch upon it here.
Parallel data means that utterances of identical linguistic content are available from both the source and the target speaker.
While this type of data makes it easier to learn a mapping of the features that capture speaker identity while keeping the content unchanged, they are harder to procure in sufficient quantities, especially for the more data-hungry \ac{DL} methods.
For this reason, algorithms relying on more non-parallel data have become more prominent in recent years.

In general, the power of \ac{DL} comes in its ability to learn complicated mapping functions from data.
In voice conversion, this ability is used to learn a transformation from an input speech signal to the target, usually by transforming the features of the source speaker to those of the target speaker before vocoding.
In deep voice conversion, this mapping can be achieved through a conditioning mechanism like the one introduced by WaveNet~\citep{Oord16-WAG}.
As discussed in the previous section, WaveNet supports both global and local conditioning -- these conditioning interfaces can be co-opted by voice conversion algorithms to change supra-segmental and segmental attributes, respectively.
The representation of speaker identity thus becomes an important aspect of \ac{VC}.
This can be done either by one-hot encodings of a fixed set of speakers~\citep{Hsu16-VCF}, d-vectors~\citep{Saito18-NVC}, or bottleneck features as speaker representations from a \ac{DNN}~\citep{LiLY0Y18}.

As with \ac{TTS}, \acp{GAN} also play a prominant role in deep voice conversions.
A \ac{GAN}-\ac{VC} framework is formulated by using the generator to map an utterance from the source to the target speaker, with the discriminator used to guide the training by classifying whether the target speaker is indeed the correct one.
As this mapping can be difficult to learn from non-parallel data, an additional form of regularisation is proposed by the introduction of a cycle-consistency loss~\citep{Zhu17-UIT}, resulting in \acs{CycleGAN}~\citep{Kaneko18-CNV, Kaneko19-CIC}.
\acs{CycleGAN} has two generators, one for transforming the speech of the source speaker to the target one, and one for the inverse conversion.
This is utilised to map the speech of the source/target speaker to the target/source one and back and ensure consistency (via the L1 loss) with the original source/target utterance.
In the process, this ensures that the wanted generator (source to target) is properly trained.
An extension of \ac{CycleGAN} for multiple speakers is found in StarGAN~\citep{Choi17-SUG, Kameoka18-SNM}, which extends the consistently principle to multiple source-target domains.
Finally, \ac{S2S} models, which use an encoder-decoder architecture, have also been extensively studied in the \ac{VC} field~\citep{TanakaKKH19, LiuCKHL00M20}.
These models have the added benefit of handling changes in sequence length induced by changes in features.
For example, a change in prosody can make the utterance of the target speaker shorter or longer than that of the source speaker, which cannot be easily handled by frame-to-frame mapping methods.

\section{Emotional speech synthesis}
\label{sec:ess}

\Acf{ESS} is that specific module of an affective agent which incorporates emotional information in speech utterances by controlling those aspects of speech which are mostly --ideally exclusively-- related to emotion.
In its broadest sense, \ac{ESS} would include a \ac{TTS} sub-component, as generating emotional speech does in fact entail the generation of non-emotional speech, meaning utterances that are comprehensible for their intended linguistic and emotional meaning.
However, most contemporary works envision \ac{ESS} (or, `expressive' \ac{TTS} as they often name it) as an extension of \ac{TTS}.
This is motivated by two pragmatic reasons: First, \ac{TTS} is a more `basic' problem than \ac{ESS}, as being unable to procure a comprehensible utterance would make any emotional fluctuations applied to it utterly meaningless.
Second, as a corollary of that fact, the research efforts placed on \ac{TTS} vastly outmatch those placed on \ac{ESS}.
As a result, most approaches are tailored to the former, and the latter is left as a mere afterthought.
For these two reasons, \ac{ESS} approaches mostly rely on an \acf{EVC} paradigm which, like \ac{VC}, consists of modifications applied to an existing \ac{TTS} model to control its emotion.
Accordingly, the majority of our review will focus on such efforts.
As for \ac{TTS}, we start with a brief history of earlier works.

\subsection{A brief history of emotional speech synthesis}
\label{subsec:early-ess}
Cahn's Affect Editor~\citep{Cahn89-AE, Cahn90-AE} and Murray's HAMLET~\citep{Murray89-HAMLET, Murray93-TTS} represent the first approaches to \acl{ESS}.
They were both rule-based and relied on the modulation of acoustic correlates of emotion (primarily pitch and timing) and providing those to existing speech synthesisers that took care of the synthesis.
The values of these parameters were usually chosen by experts, and their suitability verified by follow-up recognition studies.
A more data-driven study by \citet{Burkhardt00-VAC} investigated instead different parameter ranges and identified those that lead to better recognition rates, rather than setting them a priori, but nevertheless relied on customly manipulating these parameters for synthesis.

While rule-based \ac{ESS} brought some initial excitement to the field, it was later abandoned in favour of concatenative synthesis~\citep{Schroeder01-ESS}.
Like its \ac{TTS} counterpart, concatenative \ac{ESS} relied on selecting speech units uttered with the appropriate emotions from an existing dataset.
As a result, it too suffered from the same shortcomings that plagued standard speech synthesis, namely the lack of available data and discontinuities.

Finally, following a similar trend as \ac{TTS}, \ac{ESS} transitioned to a data-driven paradigm with the advent of \ac{SPSS}~\citep{Tachibana04-HMM, Tao06-PC}, which in turn formed a predecessor to deep \ac{ESS}.
In this context, \ac{ESS} was primarily envisioned as an intervention on acoustic features before the vocoding step: those features would be mapped to their emotional equivalents before being used to synthesise speech.
In particular, mappings between both prosodic and spectral features were learnt using data~\citep{Tachibana04-HMM, Tao06-PC}.
As is the case for \acl{VC}, this entailed the presence of \emph{parallel data} from whence the mappings can be learnt.

\subsection{Taxonomy of deep ESS approaches}
\begin{figure*}[t]
    \centering
    \includegraphics[width=\textwidth]{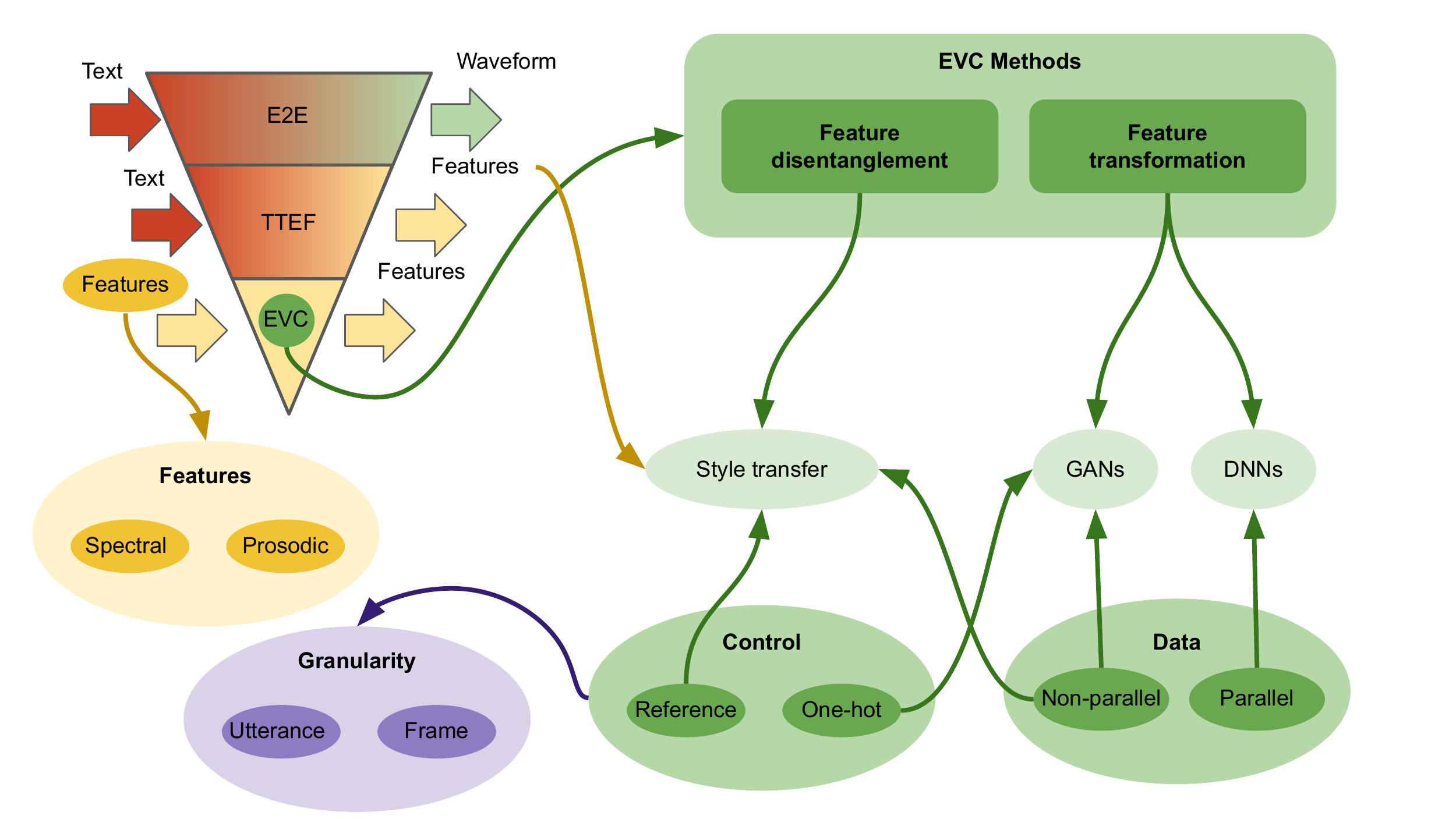}
    \caption{
    Taxonomy of deep \acl{ESS} approaches.
    Approaches can be primarily differentiated according to the following ways:
    (a) How many steps of the synthesis they incorporate, which is in term determined by their input and output, accordingly categorised as end-to-end (E2E), text-to-emotional-features (TTEF), or emotional voice conversion (EVC) methods.
    (b) How control is achieved, as well as the level of granularity that this control can achieve.
    (c) For EVC methods, on whether they use parallel or non-parallel data.
    (d) For non-parallel data EVC methods, based on whether they rely on disentangling speech components or directly mapping features to capture the target emotion, as all parallel data methods use the latter form of conversion; TTEF methods instead primarily fall under the style-transfer category.
    }
    \label{fig:taxonomy}
\end{figure*}

\begin{table*}[t]
    \centering
    \caption{
    Major \ac{DL}-based \acl{ESS} works categorised according to the taxonomy presented in \cref{fig:taxonomy}.
    }
    \label{tab:ess:taxonomised}
    \resizebox{\textwidth}{!}{
    \begin{tabular}{l c c c c c c c c}
        \toprule
        \textbf{Approach} & \textbf{Control} & \textbf{Intensity} & \textbf{Granularity} & \textbf{Non-parallel data} & \textbf{Conversion} & \textbf{Model} & \textbf{Features} & \textbf{End-to-end}\\
        \midrule
        \citet{Ming16-DBL} & Fixed & N/A & Utterance & \xmark & Transformation & bLSTM & STRAIGHT & EVC\\
        \citet{Lee17-EEN} & One-hot & N/A & Utterance & \xmark & N/A & Seq2Seq & Spectra & TTEF\\
        \citet{Lorenzo-Trueba18-IDR} & Annotator agreement & Annotator agreement & Utterance & \xmark & N/A & RNN & WORLD & TTEF\\
        \citet{Choi19-MEA} & Reference & N/A & Utterance & \xmark & Disentanglement & CNN & Spectra & TTEF\\
        \citet{Kwon19-ESS} & Reference & N/A & Utterance & \xmark & Disentanglement & Seq2Seq & Spectra & TTEF\\
        \citet{Shankar19-MSE} & Fixed & N/A & Utterance & \xmark & Transformation & Highway & F0/intensity & EVC\\
        \citet{Bao19-CES} & Fixed & N/A & Utterance & \cmark & Transformation & CycleGAN & openSMILE & EVC\\
        \citet{Luo19-EVC} & Fixed & N/A & Utterance & \xmark & Transformation & GAN & F0 & EVC\\
        \citet{Robinson19-SMO} & Fixed & N/A & Frame & \xmark & Transformation & Seq2Seq & F0 & EVC\\
        \citet{Gao19-NES} & Reference & N/A & Utterance & \cmark & Disentanglement & GAN & F0/Spectra & EVC\\
        \citet{Kim20-EVC} & Reference & N/A & Utterance & \xmark & Disentanglement & Seq2Seq & Spectra & EVC\\
        \citet{Rizos20-SFE} & One-hot & N/A & Utterance & \cmark & Transformation & StarGAN & Cepstra & EVC\\
        \citet{Cao20-VAE} & Fixed & N/A & Utterance & \cmark & Transformation & VAE-GAN & Cepstra & EVC\\
        \citet{Schnell21-EII} & Reference & Saliency maps & Frame & \xmark & Transformation & RNN & WORLD & EVC\\
        \citet{Liu21-RL} & Reference & N/A & Utterance & \cmark & Transformation & Seq2Seq & Spectra & TTEF\\
        \citet{Du21-EVC} & Reference & N/A & Utterance & \cmark & Transformation & StarGAN & Cepstra & EVC\\
        \citet{Choi21-S2S} & Reference & Manual & Utterance & \xmark & Disentanglement & Seq2Seq & Spectra & EVC\\
        \citet{Cai21-ECS} & Reference & N/A & Utterance & \cmark &Disentanglement & Seq2Seq & Spectra & TTEF\\
        \citet{Wu21-EBE} & Reference & N/A & Frame & \xmark & Disentanglement & Seq2Seq & Spectra & TTEF\\
        \citet{Kreuk21-TSE} & Fixed & N/A & Frame & \xmark & Transformation & Seq2Seq & Spectra/F0/T & EVC\\
        \citet{Zhou22-EIC} & Reference & Ranking & Utterance & \cmark & Disentanglement & Seq2Seq & Spectra & EVC\\
        \citet{Zhang22-iEmoTTS} & Reference & Posterior & Utterance & \cmark & Disentanglement & Seq2Seq & Spectra/F0 & EVC\\
        \citet{Li22-CSED} & Reference & Manual & Utterance & \cmark & Disentanglement & Seq2Seq & Spectra & TTEF\\
        \citet{Liu22-ETTS} & Reference & N/A & Frame & \cmark & Disentanglement & Seq2Seq & Spectra & TTEF\\
        \citet{Lei22-MsEmoTTS} & Reference & Ranking & Frame & \cmark & Disentanglement & Seq2Seq & Spectra & TTEF\\
        \bottomrule
    \end{tabular}
    }
\end{table*}

There are various ways in which to taxonomise deep \ac{ESS} approaches, as shown in \cref{fig:taxonomy}.
The first one is based on whether they perform \textbf{text-to-emotional-features synthesis (TTEF)}, where they go directly from text to emotionally-laden acoustic features, or \textbf{\acf{EVC}}, where they rely on acoustic features which are already generated by a standard \ac{TTS} system.
Due to the widespread success of \ac{TTS}, most \ac{ESS} systems are essentially performing \ac{EVC}, as they utilise existing components which have proven to work well.
One could argue that \ac{EVC} takes a `shortcut' compared to TTEF, since it endows an existing utterance with emotional intonation, rather than synthesising one from the ground up.
This allows \ac{EVC} to be added as an extra step in a traditional \ac{TTS} pipeline -- first synthesise, and then convert to the target emotion.
This decomposition into two constituents can reduce computational complexity and dependence on data.
Furthermore, vocoding is to our knowledge rarely explicitly adapted for \ac{ESS}, but instead existing \ac{TTS} vocoders are used out-of-the-box\footnote{One exception is \citet{Kreuk21-TSE}, which use the emotion label to condition a HiFi-GAN vocoder.}.
However, as the borders between discrete steps of a traditional \ac{SPSS} pipeline are eroding with the advent of \acl{DL}, the distinctions between these two forms of \ac{ESS} are also blurring.
Still, as most existing works go under the auspices of \ac{EVC}, these will form the majority of our review.

\ac{EVC} methods can be broken down into further categories based on the type of data they require for training.
Approaches relying on \textbf{parallel data} learn a mapping from neutral to emotional speech (or from one emotion to another) by keeping all other factors, such as the speaker or the content, constant; as this approach cannot scale well due to its strict requirements, we only touch upon them in brief.
Instead, we focus more on approaches which can work on \textbf{non-parallel data}; these scale better as the data can be pooled from several heterogeneous sources.
However, they are also more challenging, as the crucial problem of \emph{disentanglement} arises.
This challenge gives rise to the second differentiating factor: how the mapping is performed.
Some approaches choose a \textbf{direct transformation} of one type of speech to another; others opt for a \textbf{decomposition} of a speech utterance into discrete components -- emotions are one of them and thus, synthesis can be controlled by choosing one emotional `style' over another.

How emotions are represented to achieve this control is another aspect we take into account.
Here, we differentiate between \textbf{reference-based} and \textbf{reference-free} approaches.
The first kind uses an emotional speech sample to condition an \ac{ESS} system to the emotion it needs to produce.
The latter provides instead a non-auditory representation of emotions, with the choice of representation being a sub-category of differentiation.
Most previous works focus on a limited set of categorical emotions; they therefore encode them in `one-hot' vectors (`one-hot' vectors are essentially dummy variables which transform categorical labels to a numeric representation by using a vector of dimension equal to the number of categories and setting only one of its elements to 1 to represent each category);
some works also rely on `fixed' setups, where different \acp{DNN} are trained for each one-to-one mapping targeted by the \ac{ESS} system.
Reference-based methods aim for more fine-grained control: this can be achieved by transforming emotional labels to a representation space (usually learnt by \acp{DNN}) that can be used to increase the span of covered emotions.
This is also related to the type of \textbf{features} which are manipulated to achieve emotionality: while most choose to modify spectral features, some also use prosody; this in turn informs the type of vocoder which usually follows the feature conversion as it has to support the explicit control of features that are modified.
The type of features and control are also dependent on the desired level of \textbf{granularity}; utterance-level control is easier to achieve using embeddings, but frame- or word-level control requires more fine-grained approaches.

Finally, as for \ac{TTS} systems, there are different degrees in which approaches transition from a multi-step (or `cascade') \ac{SPSS} paradigm, where any one step can be implemented via a \ac{DNN}, towards an end-to-end architecture which incorporates all steps in a single model.
This \textbf{degree of `end-to-endedness'} will form our last main differentiating factor for \ac{ESS} approaches.
Naturally, as these methods typically modify an existing \ac{TTS} pipeline, they also inherit all its properties, like the type of architecture or underlying model, as discussed in \cref{sec:tts}.

We have categorised major \ac{DL}-based \ac{ESS} works according to our taxonomy in \cref{tab:ess:taxonomised}.
Whenever researchers experimented with more than one disjunct categories in their work, we chose to assign them according to where their major focus lay.
In the following subsections, we proceed to analyse the discrete categories and refer to these works in detail.

\subsection{Parallel data methods}
\label{sec:ess:parallel}

Parallel data approaches rely on datasets where the same speaker(s) have recorded a set of sentences by acting the entire set of different emotions~\citep{Lorenzo-Trueba18-IDR, Kwon19-ESS, Kim20-EVC} -- similar to parallel \ac{VC} approaches. 
This simplifies the conversion problem by keeping all other factors constant; the only thing that needs to be converted is the emotion itself.
As such, parallel data methods fall exclusively under the direct transformation category, where a \ac{DNN} is utilised to learn the mapping between acoustic features, or even directly from text and acoustic features.

However, a major downside of such methods is that they fail to scale, as collecting datasets of sufficient size is difficult given the strict requirements.
Moreover, they lack in terms of controllability.
As the mapping is dependent on the existence of parallel data, it is only possible to map from one emotion, or from neutral speech, to another type of emotion, and this mapping is fixed.
Often, it is the case that researchers train distinct networks for each combination in their dataset.

Finally, parallel datasets are often recorded in very controlled conditions.
Usually, a single speaker, or a small set of speakers, record a small set of sentences in one room using the same microphone, and are acting the required emotions.
This vastly differs from real-world situations where emotions have to be naturalistic and fit a number of different environments.
Therefore, parallel data methods cannot generalise well in the scenarios expected for real-world applications.

In conclusion, while simplifying the conversion problem by fixing other variables offers several advantages, primarily via simplifying the underlying problem, the downsides limit the applicability of the developed \ac{ESS} systems.
Thus, parallel data methods were mostly pursued in the early days of deep \ac{ESS} as a means of prototyping.

\subsection{Non-parallel data methods}
\label{sec:ess:non-parallel}

Transitioning to methods capable of handling non-parallel data was a necessary prerequisite for the development of more naturalistic and generalisable \ac{ESS} systems.
This enabled the use of larger datasets, often ones used in \ac{SER} research, collected in less controlled conditions.

This transition, though, introduced a major challenge: several factors were now entangled in the utterances to be processed.
In particular, there was now no matching sentence of the target emotion for the text that needed to be synthesised (or converted), and sometimes the data for the target emotion even came from a different speaker.
This necessitated the disentanglement of those different factors.
This entails the decomposition of an input utterance to a set of independent constituents, the modification of the emotional style (and, if needed, the speaker identity) and the reconstruction of the resulting waveform.

As we show in the next two subsections, this decomposition could be implicitly enforced to the model via manipulation of its training strategy, leading to a set of methods we name \emph{direct feature transformation methods}, or explicitly designed into it, leading to \emph{disentanglement methods}.
In either case, the increased complexity of handling non-parallel data, followed by the concurrent advancement of \ac{TTS} systems, resulted in most researchers adopting an \ac{EVC} paradigm.
They resorted to modifying the acoustic features to induce emotionality, and relied on existing \ac{TTS} pipelines for all other aspects of the synthesis process.
Only recently did they transition back to TTEF approaches, following the success of Tacotron and similar \ac{TTS} models.
This trend is also reflected in \cref{tab:ess:taxonomised}.

\subsection{Feature transformation methods}

\begin{figure*}
    \centering
    \includegraphics[width=\textwidth]{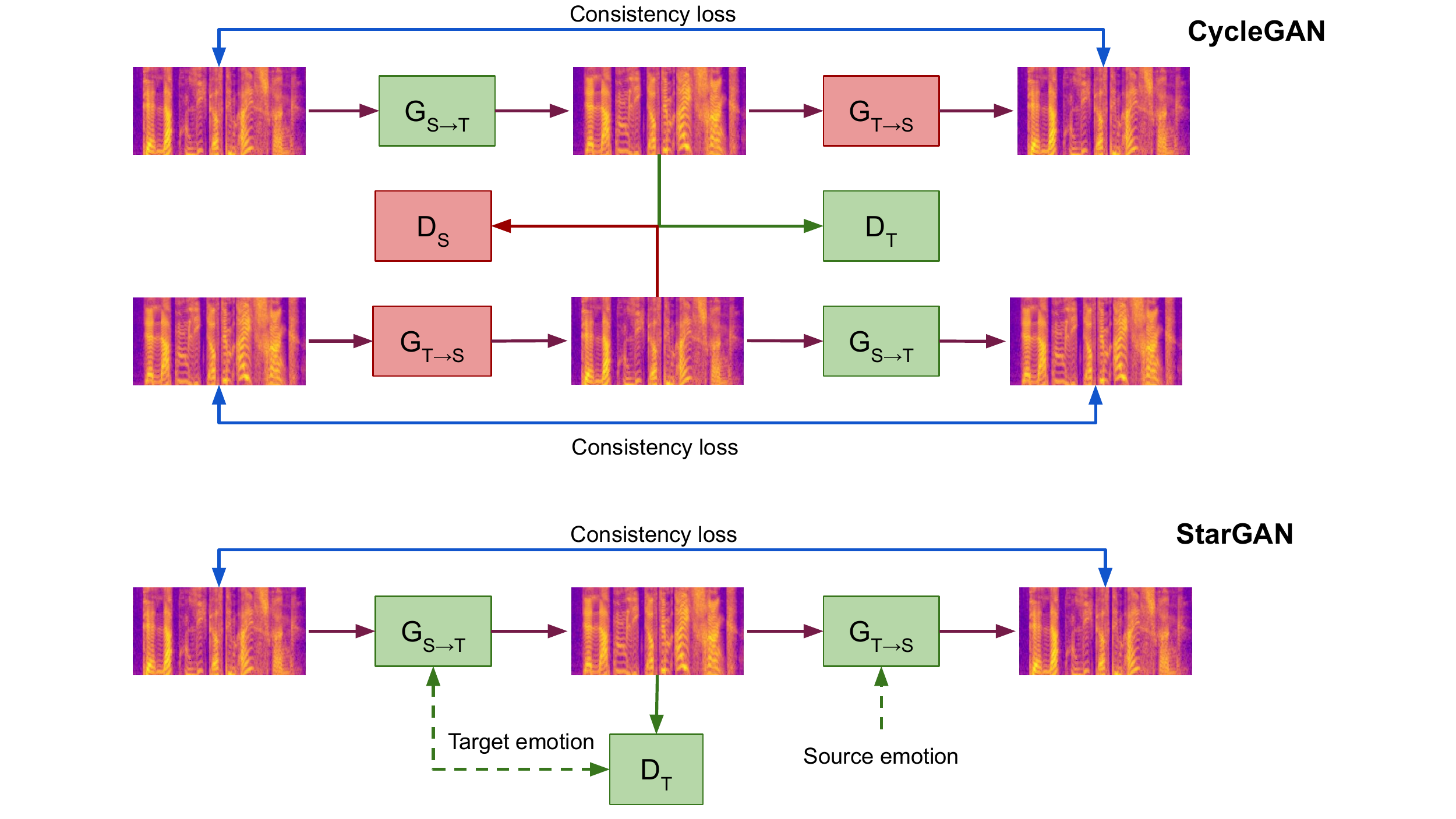}
    \caption{
    Overview of the two main \ac{GAN} paradigms for \acl{EVC}.
    CycleGAN (top) first converts the source utterance to the target domain, which is then evaluated by the discriminator, while another generator maps it back to its source domain to compute the consistency loss.
    The inverse process is followed to map from the target domain to the source domain which helps to regularise training.
    StarGAN (bottom) extends CycleGAN to handle a many-to-many mapping by training a single generator and discriminator pair, both of which can be controlled via one-hot labels to perform the correct mapping/check.
    }
    \label{fig:GAN}
\end{figure*}

The first category of \ac{EVC} methods attempts to learn a direct transformation between features of one emotion to another.
In its simplest form, the \ac{EVC} problem can be formulated as follows:
Given a set of input features $X_S \in \mathbb{R}^{T_{s} \times d}$, with $T_{S}$ being the duration of the utterance and $d$ the dimensionality of the features, the goal of an \ac{EVC} model is to map those to $X_T \in \mathbb{R}^{T_{t} \times d}$, which are the target features of a potentially different duration but of the same dimensionality.
This mapping takes the form of a function $f(\cdot): X_S \rightarrow X_T$ which will be approximated by a \ac{DNN} as $\tilde{f}(\cdot)$.
Concretely:
\begin{equation}
    x_t = f(x_s) \approx \tilde{f}(x_s)\text{,}
\end{equation}
where $x_s$ and $x_t$ are samples sourced from $X_S$ and $X_T$, respectively.
As we saw in \cref{sec:ess:parallel}, this mapping is easier to learn when $X_T$ and $X_S$ have the same linguistic content and come from the same speaker; it is then sufficient to train a \ac{DNN} to estimate the mapping between paired utterances as all this \ac{DNN} will capture is the change in emotion.
However, non-parallel methods that rely on direct transformation still have to deal with the problem of entangled factors.

The most widespread method that deals with this problem are \acfp{GAN}.
\acp{GAN} were first introduced by \citet{Goodfellow14-GAN}, and follow the basic idea of having two neural networks that compete against each other, hence the name \emph{adversarial}.
In the originally proposed \ac{GAN} framework, one of those networks, the so-called \emph{generator}, learns a function that transforms noise vectors $z$ sampled by a particular random distribution to the data $x$ that follows another distribution, where this target distribution resembles data of a specific target domain.
The second network, the \emph{discriminator}, tries to classify those artificially created instances as `fake' data.
During training, the discriminator is fed with real data from a given training set and fake data created by the generator.
Its objective is to distinguish between these two classes of data. 
On the contrary, the generator's goal is to `fool' the discriminator by learning to generate realistic data. 
Thus, the two networks have conflicting goals, resulting in both of them mutually improving each other during training.
The combined objective function $V$ of the \ac{GAN} framework can be formalised as follows:
\begin{equation}
\begin{split}
    \underset{G}{\operatorname{min}}\,\underset{D}{\operatorname{max}}\,V(D,G)= &\mathbb{E}_{x\sim p (x)}[log D(x)] \\
    &+ \mathbb{E}_{z\sim p_{z} (z)}[log (1 - D(G(z)))]\text{,}
\end{split}
\end{equation}
where $G$ is the generator network and $D$ is the discriminator.
Since the input vectors $z$ were sampled from a random probability distribution function, the generation of completely new data is possible by merely feeding different random vectors into the generator.

An intuitive extension of this principle for \ac{EVC} would be to replace the random noise vectors with data that follows the distribution of a certain speech type, \ie, a dataset of natural speech, while using the output of the generator and a dataset of the target speech type as input for the discriminator.
This would lead to the generator transforming the input speech of a certain source domain to speech of the respective target domain.
However, these kinds of translation approaches only work well with paired training data, as otherwise, the discriminator would easily detect the distributional changes induced by differences in speaker or linguistic content.
As a consequence, the standard \ac{GAN} paradigm needed to be modified for non-parallel training data.

One of the most prominent approaches that deal with the aforementioned problem is the \acf{CycleGAN}~\citep{Zhu17-UIT}.
\ac{CycleGAN} combines two unique \acp{GAN}, each consisting of its own generator and discriminator. 
The idea is that one \ac{GAN} learns to transform data from a domain $X_S$ to a domain $X_T$, whereas the other \ac{GAN} learns the exact opposite: converting data from domain $X_T$ to domain $X_S$. 
Thus, by feeding input data of one domain to one of the \acp{GAN}, and subsequently feeding the output of that first \ac{GAN} back into the second one, the final result can be compared with the original input. 
In the case of both \acp{GAN} working perfectly, the final result should be exactly the same as the initial input. 
During training, \ac{CycleGAN} uses a \emph{cycle-consistency loss} as part of its objective function, in addition to the adversarial loss that is adopted from the original GAN architecture.
The cycle-consistency loss is formulated as:
\begin{equation}
\begin{split}
\mathcal{L}_{cycle} =
&\mathbb{E}_{x_s\sim p(X_S)}[\norm{G_{T\rightarrow S}(G_{S\rightarrow T}(x_s)) - x_s}_1] \\
& + \mathbb{E}_{x_t\sim p (X_T)}[\norm{G_{S\rightarrow T}(G_{T\rightarrow S}(x_t)) - x_t}_1],
\end{split}
\end{equation}
where $G_{S\rightarrow T}$ and $G_{T\rightarrow S}$ are the generators of the two GANs, and $x_s$ and $x_t$ are data from the domains $X_S$ and $X_T$, respectively.
The full loss function of a \ac{CycleGAN} is given as:
\begin{equation}
\begin{split}
\mathcal{L} = 
&\mathcal{L}_{GAN}(G_{T\rightarrow S}, D_{S}, X_S, X_T) \\
&+ \mathcal{L}_{GAN}(G_{S\rightarrow T}, D_{T}, X_T, X_S) \\
&+ \lambda\mathcal{L}_{cycle}(G_{T\rightarrow S}, G_{S\rightarrow T}),
\end{split}
\end{equation}
where $D_{S}$ and $D_{T}$ are the discriminators of the two \acp{GAN}, and $\lambda$ a balancing factor.

\ac{CycleGAN} was first transferred to the speech domain by \citet{Kaneko18-CNV}, who used it to perform \acl{VC}.
In order to do so, they enhanced the generator networks with \emph{gated \acp{CNN}}~\citep{Dauphin17-LMG}. 
Additionally, they implemented an \emph{identity loss}~\citep{Taigman17-UCD}, which ensures that samples already belonging to the target domain are not altered. 
They formulate the identity loss as:
\begin{equation}
\begin{split}
\mathcal{L}_{identity} = 
&\mathbb{E}_{x_t\sim p (X_T)}[||G_{S\rightarrow T}(x_t) - x_t||_1] \\
& + \mathbb{E}_{x_s\sim p (X_S)}[||G_{T\rightarrow S}(x_s) - x_s||_1].
\end{split}
\end{equation}
A more sophisticated version of their work that introduces additional discriminator networks that are applied on the circularly converted voice was presented in \citet{Kaneko19-CIC}. 
Later, \ac{EVC} approaches then adopted this formulation for their purposes~\citep{Bao19-CES, Zhou20-TSA, Liu20-EVC, Fu22-ICG}.

\ac{CycleGAN} though is faced with a major limitation, namely that it only supports a translation between two domains.
However, it is desirable that \ac{EVC} methods cover a wider set of emotions.
This would mean separately training an equal set of \acp{CycleGAN} which, aside from increasing the computational overhead of experiments, also fails to benefit from the synergistic effects that may arise from a many-to-many mapping.
This problem was solved by \ac{CycleGAN}'s successor: StarGAN~\citep{Choi17-SUG, Rizos20-SFE, He21-AIS}. The training concepts of both models are illustrated in \cref{fig:GAN}.

The basic concept of StarGAN is to use a single generator and discriminator, both conditioned on features of multiple domains during training. 
The conditioning information, \ie, the emotion in the case of emotional speech conversion, is given as a domain code $c$. 
The discriminator is fed with this domain code in combination with the input audio, while the generator is conditioned with a different domain code $c'$ representing the target domain. 
Thus, the adversarial part of StarGAN's objective is formulated as follows:
\begin{equation}
\begin{split}
\mathcal{L}_{t-adv}=
&\mathbb{E}_{(x,c)\sim P(x,c)}[log D(x,c)] \\
&+ \mathbb{E}_{x\sim P(x), c'\sim P(c')} [log (1 - D(G(x,c'),c'))].
\end{split}
\end{equation}
Additionally, to enforce the model to create audio that belongs to the target domain, a classification loss is added. 
To do so, an auxiliary classifier C
is trained alongside with the discriminator and generator to distinguish between the different domains. 
This classifier is used to build the StarGAN's classification loss component:
\begin{align}
\mathcal{L}_{cls}=
\mathbb{E}_{x\sim P(x), c' \sim P(c')}[-log C(c'|G(x,c'))].
\end{align}
Furthermore, analogously to \ac{CycleGAN}, a cycle-consistency loss and an identity-mapping loss are used.

\subsection{Feature disentanglement methods}

\begin{figure*}
    \centering
    \includegraphics[width=\textwidth]{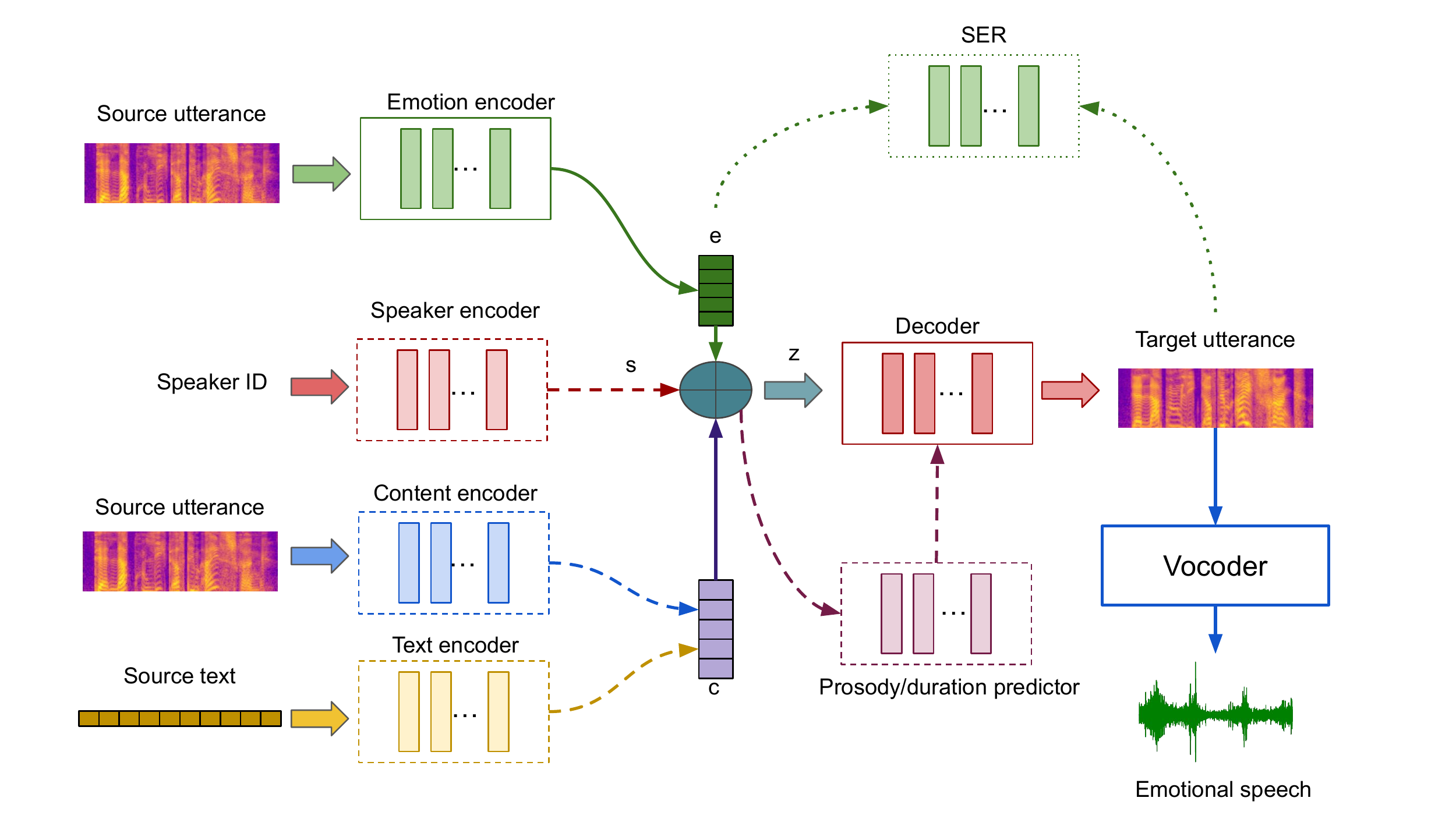}
    \caption{
    Overview of disentanglement methods for \acl{EVC}.
    The source utterance is passed through an emotion encoder to generate an emotion embedding.
    Content information is additionally provided from encoders operating on the acoustic or linguistic features -- or both.
    Optionally, speaker embeddings are provided by a speaker encoder. 
    All embeddings are fed to a decoder which generates the target utterance.
    This utterance, as well as the emotion embeddings, are often evaluated with respect to the emotion information they contain, usually by a pre-trained SER model.
    Additional information, such as prosody or duration, is often predicted from the joined embeddings and propagated to the decoder to improve the quality of synthesis.
    During inference, the source utterance or text is used to provide the content, while the reference utterance is used to generate an emotion embedding.
    }
    \label{fig:disentangle}
\end{figure*}

Disentanglement methods attempt to explicitly decompose into several constituents: emotion, linguistic content, and, potentially, speaker effects.
This is tantamount to assuming that the speech signal results from an equivalent number of latent factors, $c$ (for content), $e$ (for emotion), and $s$ (for speaker), which together form the latent code $z$ defined over a representation space $\mathbb{Z}$.
This latent code is mapped to the observation space $X \in \mathbb{R}^{T \times d}$ via a function $f : \mathbb{Z} \rightarrow \mathbb{R}^{T \times d} $.
The core idea is to preserve the factors related to content and speaker while manipulating the one related to emotion.

The basic schema, as followed by recent works~\citep{Cai21-ECS, Zhou22-EIC, Zhang22-iEmoTTS}, is shown in \cref{fig:disentangle}.
The source utterance features are passed to an encoder whose goal is to learn the content, resulting in an embedding $c_s$; a reference embedding (or some other representation of emotion) is passed to an emotion encoder, which generates an emotion embedding $e_t$; both embeddings are passed to a decoder which attempts to reconstruct the target utterance.
If no \acl{VC} is required, it is assumed that the content embedding $c_s$ also contains speaker information, which should be preserved; otherwise, a speaker embedding $s_t$ is also created by a speaker encoder. 
In order to guide the respective encoders to properly disentangle the information they need, specific losses are introduced.
For example, the embeddings of the emotion encoder could be compared to embeddings from an \ac{SER} model, or passed to an \ac{SER} model themselves such that they learn to classify the target emotion~\citep{Zhou22-EIC, Li22-CSED, Liu22-ETTS}.
This way, they are guided to learn emotional information.

        
Inspired by advances in \ac{TTS}, with models like Tacotron achieving impressive synthesis results, recent works have transitioned to conditioning such models on emotional information~\citep{Cai21-ECS, Wu21-EBE, Li22-CSED, Liu22-ETTS, Lei22-MsEmoTTS}, thus moving a step up from \ac{EVC} methods in the end-to-end hierarchy and going directly from text to audio features.
These methods have no content encoder relying on audio inputs; instead, the text encoder acts as the content encoder to capture linguistic style.
While the information fed to the content encoder is already decoupled from its expressivity (as it comes in the form of characters or phonemes), we consider these methods to be relying on disentanglement, as they still rely on an emotion encoder to capture emotional information and a content encoder to capture linguistics.
The main difference is that this decomposition is now implicit for the content part.

\subsection{Controlling the synthesised emotion}

Controlling the generated emotion is a fundamental aspect of \ac{ESS} systems.
In fact, this is the most important differentiating factor between methods falling under the \ac{ESS} umbrella and \emph{expressive \ac{TTS}} systems, which impart prosodic fluctuation on the generated signals.
Without explicit control, these fluctuations influence other aspects of the output utterance besides emotion, such as speaker identity or overall speaking `style' (for example, read vs spontaneous speech).
While \ac{ESS} methods make great efforts to properly disentangle these aspects, they also need a mechanism to control the emotion during inference.

Early \ac{ESS} methods relied on a trivial control mechanism: by independently training conversion systems for all possible emotion pairs, inference was simply done by selecting the appropriate pair.
This is essentially superseded by methods which use one-hot encodings of emotions, like StarGAN~\citep{Rizos20-SFE}; there each emotion is represented by a one-hot vector.
These approaches are better able to handle many-to-many mappings by jointly learning from several emotion pairs.

However, relying on fixed codes is far below the level of control required for successful \ac{HCI} applications.
For this reason, later methods resort to reference-based `style transfer' methods, which are inspired from recent advances in expressive \ac{TTS}.
The core proposition behind style transfer is to learn an encoding for those aspects of speech that correspond to emotion in a data-driven way.
This idea was first introduced in \citet{skerry2018towards}.
The authors utilised a reference encoder to capture prosodic variations in the reference signal and transfer them to the target utterance.
This encoder was jointly trained with the main \ac{TTS} system; during training, the reference was identical to the target utterance so a mapping could be learnt; during inference, the reference was chosen arbitrarily to impose a specific style on the target utterance.
Their reference encoder was a \ac{CNN} followed by a \ac{GRU} that relied on Mel-spectrogram features; the style embedding was simply the last state of the \ac{GRU} which was then fed to the decoder.
However, according to the authors' experiments, the learnt embeddings captured heavily entangled information, which is highly undesirable as it inhibits the fine-grained control of the synthesis process.
For example, transferring from a female voice to a male voice, resulted in an overall lower pitch, which sounded like a female trying to imitate a deeper voice~\citep{skerry2018towards}.

To further promote disentanglement between different factors, \citet{wang2018style} introduced \acp{GST}.
Utilising a similar setup as \citet{skerry2018towards}, they introduced an additional constraint on the reference encoder.
Rather than propagating the last state of the \ac{GRU} to the decoder, they first use it as the query in attention operation over a set of learnable tokens: the \acp{GST}.
These tokens, whose number was fixed a priori to $10$, would be the knobs to be twisted during inference time.
During training, the attention mechanism would `softly' weigh the contribution of each token to the reference; the contributions would then be combined and fed to the decoder.
During inference, the user can either provide a different reference, which would be accordingly ran through the reference encoder to generate the \ac{GST} weights, or directly manipulate the weights themselves to achieve the required outcome.
Inducing this constraint resulted in more naturalistic control and better disentanglement in the styles.

Both these procedures are often used for \ac{EVC}.
The main difference is in the type of utterances provided to the reference encoder.
During inference, these are selected to belong to the target emotion, similar to the expressive \ac{TTS} case where a reference is picked to fit the required style.
However, in \ac{EVC}, it is also to explicitly guide the tokens to encode the reference emotion during training~\citep{Wu19-EES, Li22-CSED, Cai21-ECS, Liu22-ETTS}.
This already biases the reference encoder to capture the differences in emotion, and results in better controllability during inference as well.
Moreover, it is often the case that an average style is computed on all training data and stored as reference for each category, therefore sparing the need to select an appropriate reference during inference.

\subsection{Controlling emotional intensity}

Another vital aspect of control is that of emotional intensity.
Emotions come not in discrete states, but in continua which define fluid categories that seamlessly transition from one to the other~\citep{Cowen19-BHV}.
As such, controlling the strength of a required emotion --over a continuous axis-- is of fundamental importance for \ac{ESS}.
In general, this area has remained relatively underexplored as researchers grapple with the challenges of discrete \ac{ESS}.
Nevertheless, there have been important advances in recent years.

\citet{Lorenzo-Trueba18-IDR} were probably the first to investigate it for deep \ac{ESS}.
They used an annotator-driven representation of emotion, which assigned a relative value to each utterance.
This relative value was computed using the confusion between the expected emotion (\ie, what the actors in their dataset were supposed to act) and the annotated emotion (\ie, what annotators perceived).
This allowed them to represent emotions via continuous, rather than discrete, vectors, which in turn allowed for the fine-grained control of emotional synthesis via manually setting those weights during inference.
A similarly manual setting was also explored by \citet{Choi21-S2S, Li22-CSED}.

While simple in its conception, this form of control is fundamentally limited by the lack of data or the need to manually tune parameters.
An interesting alternative is found in \citet{Schnell21-EII}, who use the saliency maps of pretrained, attention-based \ac{SER} systems as guidance for frame-level intensity control.
Similarly, \citet{Zhang22-iEmoTTS} use the posterior probabilities of an \ac{SER} for utterance-level emotional control.
Both approaches follow a reference-based paradigm for inference-time control.
Ultimately though, those approaches are limited by the effectiveness of the \ac{SER} systems, which, although greatly improving in recent years~\citep{Wagner22-DTE}, is still far from perfect.
Moreover, it is not necessarily the case that those references evaluated by an \ac{SER} system as more `probable' are necessarily those of a higher intensity; they could merely be those that are closest to its training distribution.

A solution is given through exploiting the inherently ordinal nature of emotions~\citep{Yannakakis17-ONE}.
\citet{Zhou22-EIC, Lei22-MsEmoTTS} exploit this fact by learning a ranking function for the intensity of each emotion.
Their approach relies on assuming that all neutral samples have an intensity of zero, and proceeding to generate emotional-neutral ranking pairs, as well as neutral-neutral and emotional-emotional anchor pairs, over which a max-margin optimisation problem is approximated.
This results in a weighting matrix $W$ which provides a ranking between $[0-1]$ for each feature vector $x$.
During inference, this ranking can be manually set to control the intensity of the synthesised emotion.

\subsection{Granularity of emotional control}
\label{ssec:granularity}
Most works impose a single emotional category or style on an entire utterance, assuming that this will be accordingly `distributed' by the mapping network or decoder to the appropriate frames.
However, achieving a more fine-grained level of control can help increase the naturalness of expressed emotions, as well as add the capacity to express more nuanced emotional states.
To that end, some works pursue more granular representations of emotion.
For example, \citet{Schnell21-EII} are able to achieve this through their saliency maps, which assign a level of control to each frame via an attention-based \ac{SER} model.
Similarly, \citet{Wu21-EBE} achieve this with a capsule network~\citep{Sabour17-DRC} while \citet{Kreuk21-TSE, Liu22-ETTS, Lei22-MsEmoTTS} achieve this via frame-level losses.
As seen in \cref{tab:ess:taxonomised}, this trend is picking up pace this last year with several very recent works pursuing higher degrees of granularity.

\subsection{Features manipulated to achieve expressivity}
\label{ssec:features}

As \cref{tab:ess:taxonomised} shows, most methods fall under the \ac{EVC} category, meaning that they primarily manipulate acoustic features to achieve expressivity.
These features are to a large extent motivated by the decades of research devoted to understanding which facets of speech are impacted by emotion and how.
This research is touched upon in \cref{ssec:correlates}.
Another factor which influenced the choice of features is the success of modern \ac{TTS} architectures.
As we saw, \ac{EVC} methods heavily rely, and often outright incorporate, existing \ac{TTS} pipelines.
It is only natural that they then use the same features that those \ac{TTS} pipelines support.
Previously, this restricted the set of features to ones supported by \ac{SPSS} vocoders, such as WORLD or STRAIGHT, which included F0, spectral/cepstral, and aperiodicity features~\citep{Ming16-DBL, Lorenzo-Trueba18-IDR, Robinson19-SMO, Rizos20-SFE, Schnell21-EII}, while works that only intended to use \ac{EVC} as a means to improve \ac{SER} performance~\citep{Bao19-CES} manipulated feature vectors used by the downstream models~\citep{Eyben10-openSMILE}.
Nowadays, with the advances seen in neural vocoders, it is typical to use those as the last step of the synthesis process; accordingly, \ac{EVC} pipelines now concentrate more on modulating spectral features~\citep{Lee17-EEN, Choi19-MEA, Kwon19-ESS, Kim20-EVC, Liu20-EVC, Zhou22-EIC}.
Manipulation of other features, such as F0, is still done but primarily using simple statistical techniques (\eg, by standardising the F0 curve with the statistics of the target emotions~\citep{Rizos20-SFE}) and rarely using deep learning methods~\citep{Shankar19-MSE, Kreuk21-TSE}.
Overall, this shows that the field is transitioning to a standard of using more abstract representations (spectrograms) and relying on the representation power of \acp{DNN} for learning to modify the appropriate signal characteristics.

\subsection{Deep models used in ESS research}

In general, from an architecture perspective, the innovation in the field of \ac{ESS} does not seem targeted to novel \ac{DL} models, but rather on finding novel ways of combining existing modules to achieve desired effects (\eg, disentanglement).
This is to be expected following the success of \ac{TTS}; adopting best-practices from a neighbouring field allows the community to iterate quickly over problems that are specific to \ac{ESS} rather than reinventing the wheel.
As seen from \cref{tab:ess:taxonomised}, the majority of \ac{EVC} models are relying on \ac{S2S} models~\citep{Lee17-EEN, Kwon19-ESS, Robinson19-SMO, Kim20-EVC, Schnell21-EII, Liu21-RL, Zhou22-EIC, Zhang22-iEmoTTS}.
This is counter to earlier methods, which relied on highway networks or simpler sequential models~\citep{Ming16-DBL, Lorenzo-Trueba18-IDR, Choi19-MEA, Shankar19-MSE}.
The main downside of those was that they could not handle the differences in signal duration that resulted from a change of emotion; thus, this mapping of duration needed to be handled explicitly.
In contrast, \ac{S2S} methods have a natural way of handling the change in duration as the decoder can reconstruct sequences of different length than those seen by the encoder~\citep{Yang22-S2S}.
A more thorough overview of \ac{S2S} models for \ac{EVC} can be found in \citet{Yang22-S2S}.
This \ac{S2S} trend is also followed by more recent TTEF methods, which rely on the Tacotron architecture~\citep{Cai21-ECS, Wu21-EBE, Li22-CSED, Liu22-ETTS, Lei22-MsEmoTTS} and merely condition it with emotional information.
Methods using adversarial models like CycleGAN~\citep{Bao19-CES} and StarGAN~\citep{Rizos20-SFE} also stay close to their original versions, with minor adaptations to fit the \ac{EVC} problem.
Finally, some works attempt to leverage representations learnt by large, pre-trained models and thus rely on transformer-based architectures such as HuBERT~\citep{Kreuk21-TSE}.

The decomposition of input utterances, either source or reference, is typically achieved via the use of \acp{AE}; this makes them foundational building blocks of several \ac{EVC} methods; they therefore warrant a closer analysis.
Traditional \acp{AE} are comprised of two parts: (a) an encoder, which reduces the dimensions of the speech signal to a latent representation (or code), and (b) a decoder, which tries to reconstruct the original speech representation from the code. 
Mathematically, given the speech frame of dimension $d$ $\{x \in R^{d}\}$, we define the encoder as a function $q: \mathcal{X} \rightarrow \mathcal{Z}$ such that $z = q_{\phi}(x)$ with parameters $\phi$, and the decoder as a function: $\psi: \mathcal{Z} \rightarrow \mathcal{X}$, such that $\tilde{x} = \psi_{\theta}(z)$ with parameters $\theta$.
The same principles are used for \ac{EVC}; however, instead of a single encoder there are often multiple ones, one for each latent factor that needs to be disentangled, while a single decoder takes care of the inverse mapping to the feature space.

A probabilistic realisation of \ac{AE} that is sometimes used for \ac{EVC} as well is the \ac{VAE}~\citep{Cao20-VAE}. 
A \ac{VAE} is used to generate the speech representation of the target domain, where the code of the network is assumed to be represented by a Gaussian distribution $\textbf{z}\mathtt{\sim}\mathcal{N}(\textbf{z};\mathbf{\mu}, diag(\sigma^2))$. 
The encoder tries to estimate the mean and variance of the distribution, and with the use of the reparameterisation trick we can sample a code representation. 
The code is fed to the decoder, which estimates a new speech representation. 
The training is performed by maximising a variational lower bound of the log-likelihood:
\begin{equation}
    \mathcal{L}( \theta, \phi; x) = -D_{KL} (q_{\phi}(z | x) || p(z)) + E_{q_{\phi}} [\log \psi_{\theta} (x|z)],
\end{equation}
\noindent
where $D_{KL}(q||\psi)$ denotes the Kullback-Leibler divergence between the distributions $q$ and $p$.

\subsection{Evaluation protocols}
\label{ssec:evaluation}

Protagoras of Abdera famously claimed that ``Of all things the measure is Man, of the things that are, that they are, and of the things that are not, that they are not''.
So is the case for the evaluation of \ac{ESS} approaches as well, with the employment of human annotators being the gold standard for judging the effectiveness of \ac{ESS} approaches.
The most commonly used process is a judgement test, where annotators are asked to evaluate the similarity of a generated signal with respect to a reference stimulus, or, in the reference-free variant of those tests, to simply classify the emotion of the generated signal.
Alternatively, they are asked to evaluate different signals with respect to different aspects that correspond to emotional speech, such as likeability, emotional strength, or naturalness.
In all cases, individual ratings are aggregated to procure a final \acf{MOS}~\citep{Streijl16-MOS}.
Usually, these ratings are on a $[0, 5]$ scale with steps of $0.5$, with $5$ being the best score.
While there is currently a dearth of well-established dimensions on which to evaluate emotional speech, the field is drawing inspiration from the much more mature metrics for \ac{TTS}~\citep{Hinterleitner17-AAI}.
Some dimensions commonly used in recent works are naturalness~\citep{Luo19-EVC, Gao19-NES, Choi21-S2S, Liu22-ETTS}, speech quality~\citep{Schnell21-EII, Du21-EVC}, emotional strength~\citep{Lorenzo-Trueba18-IDR}, and similarity with the target emotion~\citep{Choi19-MEA, Kwon19-ESS, Wu21-EBE, Zhang22-iEmoTTS, Li22-CSED}.
ABX tests are also commonly used~\citep{Luo19-EVC, Du21-EVC}, where subjects are asked to tell if sample X, which is randomly chosen from category A or category B, is closer to a reference from A or a reference from B~\citep{Huang98-ABX}.
If subjects systematically pick the correct category for sample X, the two categories are considered to be distinct enough.
Finally, some works evaluate \ac{ESS} approaches by how well annotators are able to classify the synthesised emotions~\citep{Robinson19-SMO, Cao20-VAE, Schnell21-EII, Liu21-RL, Choi21-S2S, Cai21-ECS}.
        
As human evaluations are often costly and time-consuming, the community has attempted to supplement them with automatic ones.
These evaluations are based on algorithmic measures that quantify different signal properties~\citep{Hinterleitner17-AAI}.
In the case of \ac{TTS}, for example, BSD~\citep{Wang92-AOM}, PESQ~\citep{Rix01-PEO}, POLQA~\citep{Beerends13-POL}, or ITU-T Rec. P.563~\citep{Malfait06-TIS} are often used to evaluate the quality of generated signals.
No such standardised procedure exists yet for \ac{ESS}, but several researchers are using distance metrics (\eg, Euclidean) between generated and target features, such as Mel spectra, or even using pre-trained \ac{SER} models to judge whether generated samples are correctly classified~\citep{Baird21-PNA}.
These metrics, though far from error-free, vastly speed up the development process of \ac{ESS} approaches by providing quick feedback to researchers, and are thus an integral part of the \ac{ESS} ecosystem.

\subsection{Datasets of emotional speech}
In a data-driven paradigm, datasets become the cornerstone of successful models.
A comprehensive overview of existing datasets of emotional speech used in \ac{ESS} can be found in \citet{Zhou22-ESD}.
The authors mention five key desiderata for designing datasets that cover all conditions necessary for generalisation:
\begin{enumerate*}
    \item Increasing lexical variability, as emotional speech datasets are often recorded using a limited set of datasets.
    \item Introducing language variability, as \ac{ESS} approaches might be expected to work for different languages and cultures.
    \item Promoting speaker variability, as acted datasets are typically recorded from a few actors and thus do not generalise well to new speakers.
    \item Controlling for confounders, such as different accents or demographics.
    \item Regulating recording conditions, both to control unwanted confounders and to safeguard the quality of ground truth samples. 
\end{enumerate*}
However, this last factor can also act as an inhibiting factor for \ac{ESS} applications that should generalise to different background environments; thus, we consider it a good restriction while the field is still in its nascent stages, but one that must ultimately be abandoned as we transition to more realistic applications.
The authors also introduce a new dataset, ESD, which is now being increasingly used by the community as a standard benchmark.
Prior to the introduction of ESD, researchers used either small scale datasets created explicitly for \ac{ESS}~\citep{Lorenzo-Trueba18-IDR, Kwon19-ESS, Kim20-EVC} or relied on the standard \ac{SER} datasets, such as IEMOCAP~\citep{Busso08-IIE}, 
EMO-DB~\citep{Burkhardt05-EMODB}, etc.

\subsection{State-of-the-art performance}
At the end of our overview of modern \acl{ESS} approaches, one important question remains open: Is \ac{ESS} a solved problem?
Recent works boasting average emotion similarity \ac{MOS} scores of 4 for 6~\citep{Li22-CSED, Lei22-MsEmoTTS} or 7~\citep{Zhang22-iEmoTTS} emotion categories certainly suggest that we are approaching a `WaveNet moment' for \ac{ESS} as well, as the revolution started by \citet{Oord16-WAG} began with such \ac{MOS} scores for naturalness.
Accordingly, some works are showing subjective emotion recognition accuracies reaching up to 80\%~\citep{Liu21-RL, Choi21-S2S}.
Other works, however, feature much lower scores, dropping down to almost 50\% recognition accuracy for 4 emotions~\citep{Gao19-NES, Cao20-VAE}.
While one could easily dismiss the low-performing approaches as simply inferior, a closer look at the data used in each work reveals a more nuanced interpretation: \citep{Li22-CSED, Lei22-MsEmoTTS} used read speech datasets recorded by single, female authors specifically constructed for \ac{ESS}, while \citep{Gao19-NES, Cao20-VAE} both used IEMOCAP~\citep{Busso08-IIE} which includes improvised emotional speech.

This begs the question: How do we define success?
This brings us back to the original question of what makes an affective agent.
Success depends on the type of agent and the environment they are expect to operate in.
It depends on the number and kind of emotions the agent is expected to support, the languages and cultures it needs to cover, their malleability to user input, their robustness to different noise conditions, etc.
As \ac{ESS} make their journey out of research labs and into the real world, we expect fluctuations between periods of high performance on restricted conditions, followed by low valleys of \ac{MOS} scores as the application field is expanded and evaluation criteria get increasingly stricter.
Existing works show that the barrier of single-/few-speaker \ac{ESS} systems with limited acted emotions on read speech has been breached, but we are only now approaching the frontier of naturalistic emotions, as all recent works are still relying on acted emotional data.

\section{Discussion}
Our overview has shown that \acl{ESS} is a rapidly growing field which is being heavily influenced by the \acl{DL} era of \ac{AI}.
In the last subsection, we argued that while \ac{DL} constitutes an immense leap forward compared to previous approaches, \ac{ESS} remains far from solved.
In this section, we highlight the main limitations, discuss whether the \ac{ESS} problem should be solved at all given the ethical considerations it raises, and finally outline some promising areas of future research.

\subsection{Main limitations}

As seen in \cref{ssec:evaluation}, \ac{ESS} is still plagued by a lack of holistic, standardised evaluation protocols.
In particular, there is a poignant lack of automatic evaluation benchmarks that allow a fair comparison of different approaches.
As seen in other fields of \ac{AI}, benchmarks become the driving force which guides new advances.
In contrast, even though significant progress has been made in recent years in \ac{ESS}, this progress is hard to distill in a single leaderboard which highlights the most promising future directions.
More importantly, for any new algorithm that needs to be compared with the state-of-the-art, researchers have to revert to costly human evaluations that hinder the rapid advance of the field.
This makes it harder to iterate over new ideas and ascertain the impact of a proposed improvement.
It is, however, a problem that can be easily solved by a focused effort of the community to use similar datasets and report similar metrics.

A more serious challenge is achieving the amount of controllability required by downstream applications.
Disentanglement of all confounding factors that influence a speech utterance remains the `holy grail' of \acl{ESS} (and, for that matter, analysis too).
Without proper disentanglement, \ac{ESS} methods will be unable to yield a suitable set of `knobs' that an end-user can twist to generate the appropriate emotion.
This problem also plagues state-of-the-art \ac{SER} architectures~\citep{Wagner22-DTE}, where models learn an entangled representation of linguistics and acoustics~\citep{Triantafyllopoulos22-PSE}.
As \ac{ESS} is scaled up to naturalistic datasets with a bigger lexical variability, we expect this issue to arise there as well.

Overrepresentation of a few `dominant' cultures and languages is another problem; while it is motivated by pragmatic reasons, namely the availability of data, it nevertheless limits the applicability of the developed approaches.
While research in related fields, such as \ac{ASR}, shows that algorithms will generalise well to new languages once trained with data from those languages, it remains a challenge to procure data of such quantity for most of those.
The use of more data-efficient methods to drastically cut down on the demand for data is still an open issue in the \acl{DL} era of \ac{AI}, though we expect advances in neighbouring fields to trickle down to \ac{ESS} as well.
However, this overall lack of cultural representation also raises ethical concerns as to whether \ac{ESS} research can be universally applied and thus should be seriously considered by the community besides the point of finding the data (see also \cref{ssec:ethics}).

Finally, we would be amiss not to point out the fact that contemporary affective computing research shies away from the problem of endowing machines with the capacity to have emotions.
Thus, \ac{ESS} approaches adhere to the ``fake it until you make it'' mantra, whereby \ac{HCI} agents simulate the presence of emotions by appropriately modulating their voice.
However, as research in human emotions has shown, there is a noticeable difference between acted and natural emotions (which can only, if ever, be circumvented by the best of actors)~\citep{Cowie11-IDC}.
Therefore, it could be that the gap between humans and machines cannot be bridged until the latter also acquire the ability to simulate realistic emotions.



\subsection{Ethical considerations}
\label{ssec:ethics}

In recent years, it has become increasingly evident that just because \acl{AI} methods \emph{can} do something, it does not necessarily mean that they \emph{should} do it.
This is also a central question in the field of \acl{ESS}.
While the potential to dramatically improve \acl{HCI}, assist speaking-impaired individuals, and give voices to the intelligent agents of tomorrow is thrilling, there are several societal challenges facing our community in the here and now.

The most poignant of those issue is the rise of `deep fakes' (AI-fabricated videos of people saying or doing something that they have never said or done in real life)~\citep{Chesney19-DF}.
With the rapid advances in \acl{ESS}, it is not far-fetched to assume that future `deep fakes' are not only going to change the linguistic content of targeted speakers, but also their emotional one.
This vastly increases the capabilities of malignant actors to spread disinformation about, or defame, a particular individual, even without changing their choice of words.
For example, simply changing the tone of a politician who refers to a particular demographic group to sound sarcastic or derogatory could incur substantial damage to their public image.

A similar, more insidious approach would be to adapt the perceived personality of the target speaker.
This can be used to make a particular candidate more or less appealing, or even to cast a whole demographic in a particular light, by manipulating the personalities of its spokespeople, \eg, to be seen as more aggressive or submissive.
One particular example is that of voice assistants: as criticised in recent a UNESCO report~\citep{West19D}, the initial design of several voice assistants was to show submissiveness, even in the face of blatant abuse, reinforcing notions of outdated `female servility'.
This case study shows how biases can be perpetuated through technological products in particular when those relate to a simulation of behaviour and personality.
This potential to transform public opinion through the use of targeted misinformation represents a major threat to societies around the world, and would be vastly exacerbated by the improvement of \ac{ESS} algorithms.

A final aspect of whether we `should' do \ac{ESS}, is whether we want conversational agents to be emotional.
This will give them the unprecedented capability to influence our own emotions, perhaps in ways we would prefer to avoid.
For example, agents whose purpose is to elicit more sales, could adapt their voice to appear more trustworthy or friendly, thus subverting the buyer's will.
Moreover, a related question is whether artificial agents should be clearly distinguishable from humans; the EU White Paper on Artificial Intelligence explicitly states that humans should be made cognisant of the fact that they are interacting with an artificial entity under all circumstances~\citep{european2020white}, but the question remains if that is sufficient to mitigate the potential dangers that could arise from `overhumanising' those entities.

The question of `should' does, however, not cover the degree in which we `can'.
As is evident from the approaches presented here, full-blown conversational agents with the capacity to accurately and naturally convey emotion are increasingly on the way.
Still, there are still a lot of critical considerations to be addressed.
The first one is \emph{generalisability}: Do we cover all different cultures? Do we accurately represent all individuals?
The second one is \emph{privacy}: Whence do we source our data from?
The third one is \emph{correctness}: Is our evaluation sufficient?

None of these questions can be answered satisfactorily (yet).
Research in \ac{ESS} is being targeted to a small number of languages and cultures, the ones typically available in existing datasets, such as English or Chinese.
Moreover, the emotions in these datasets, and the corresponding synthesised samples, are typically annotated from  individuals of particular demographics (often students in the case of University research).
This calls into question whether we are accurately capturing all the nuances of emotional speech across different cultures.

Emotions are also one of the most precious aspects of human experiences.
Sourcing the vast quantities of those required by contemporary approaches is challenging without violating privacy.
In particular, collecting negative emotions in realistic scenarios requires us to infringe on the most private moments of an individual, such as the heartbreaking loss of a loved one.
Acted data can only get us part of the way there, but how we take the next step needed for naturalistic emotions remains an open, and challenging, question.

Evaluation is perhaps the easiest of the three questions.
Decades of research on the perception of emotional speech provides a solid background from which to start.
Co-opting those approaches for the evaluation of synthesised emotional speech, and adopting best-practices from the sister-domain of \ac{TTS}, seems like a realistic goal.

Overall, it seems obvious that \ac{ESS} leads to very serious ethical, legal, and social impact (ELSI) challenges. 
A full consideration of ELSI aspects cannot be given here, as it is too wide in scope for a transformative technology like \ac{ESS}. 
However, specifically for the field of computational paralinguistics, the reader is referred to \citet{Batliner20-EAG}.

\subsection{Future perspectives}
\label{ssec:future}
After the tremendous advances that the \ac{TTS} field saw in the last few years, \ac{ESS} seem poised to become next frontier for the speech synthesis community.
Aside from tackling existing challenges and addressing the ethical considerations raised in the previous sections, we expect a few methodological advances to capture the interest of the community.

Synthesising emotional vocal bursts is one of them.
In the now famous promotional video for Google Assistant\footnote{https://www.youtube.com/watch?v=yDI5oVn0RgM}, the crowd erupted in cheers as the assistant assured the hairdresser that ``taking one second'' to look for an appointment was fine with a mere ``Mm-hmm''.
This illustrates how vocal bursts are essential components of emotional responses~\citep{Cowen19-BHV}.
Synthesising them was already the topic of the 2020 \ac{ExVo} Challenge\footnote{https://www.competitions.hume.ai/exvo2022}.
The best-performing approach, which used StyleGAN2, already achieved promising results that highlight the potential of this line of research~\citep{Jiralerspong22-EXVO}.

Similarly, as stated in our introduction, conveying emotions is but one aspect of an affective agent.
Endowing the agent with an artificial personality is another area which has been pursued for several decades~\citep{Brown73-POP}.
This topic has been recently revived in the context of big language models, which can be adapted to emulate a specific personality~\citep{Key19-MBP}.
As personality has been also shown to manifest in speech signals~\citep{Schuller12-COMPARE}, it is an evident next step to introduce it to conversational agents as well~\citep{Andre99-IMP}.
In general, as exemplified by the tasks featured in the Computational Paralinguistics Challenge\footnote{www.compare.openaudio.eu}, there exist a plethora of speaker states and traits which can be modelled from speech: deception, sincerity, nativeness, cognitive load, likability, interest, and others are all variables which could be added to the capabilities of affective agents.

Personalisation is expected to be another major aspect of future \ac{ESS} systems.
Both the expression~\citep{Ritschel19-PSI, Baird19-DGA, Triantafyllopoulos21-DSC} and the perception~\citep{Ando21-LAM} of emotion show individualistic effects which are currently underexploited in the \ac{ESS} field.
Future approaches can benefit a lot from adopting a similar mindset and adapt the production of emotional speech to a style that fits both the speaker and the listener.
Such an interpersonal adaptation effect is also seen in human conversations and is a necessary step to foster communication~\citep{Amiriparian19-SIS}

Finally, as future affective agents find their way out of their academic research sandboxes and into the real world, they will be forced to interact with other entities -- artificial and human alike.
This will form a natural breeding grounds for interactions, which can be accordingly classified as `successful' or not, depending on the goals of the agent.
Coupled with effective \ac{SER} capabilities, these interactions constitute a natural \emph{reward signal} which can be further utilised by their agent to improve their \ac{ESS} and \ac{SER} capacities in a lifelong reinforcement learning setup, which still remains an elusive goal for the field of affective computing~\citep{Schuller21-FIVE}.
An overture to this exciting domain can already be found in intelligent dialogue generation, where reinforcement learning is already being used to adjust the linguistic style of an agent  \cite{ritschel:et:al:2017}  or to learn backchanneling responses~\citep{Bayramouglu21-ERA, Hussain22-TSE}.
We expect this paradigm to be more widely used in \ac{ESS} in the near future.

\section{Conclusion}
\label{sec:conclusion}

We have presented an overview of recent advances in the synthesis of affective speech
, including affective voice conversion. 
Deep learning is paving the way for considerable advances in this field and laying the foundation for the affective conversational agents of tomorrow.
Most work has focused on categorical emotions, using, in particular, acted datasets of read speech.
The community has mostly concentrated on modifying acoustic features, a form of emotional voice conversion, but there is recently a renaissance of \ac{ESS} approaches that directly map text to acoustics.
Accordingly, we are seeing an increasing consolidation of advances in \ac{TTS} and a move towards more `end-to-end' emotional synthesis.
Finally, following recent successes on conversion of one emotion category to another, albeit in the restricted domain of acted and read speech, several works are now focusing on the control of emotional intensity, thus increasing the controllability of EVC methods.

As main challenges to existing approaches, we have identified the absence of naturalistic emotions in the most widely-used corpora, the overrepresentation of a few cultures and languages in emotional datasets, the issue of disentangling the different latent factors that influence speech, and the inherent limitations of an approach that tries to imitate, rather than simulate emotions.
Another major challenge is the adherence to ethical rules, as machines that can simulate affect in all its manifestations, such as emotion and personality, can pose serious threats to societies in the era of `fake news'.
Nevertheless, we believe that concentrated efforts by the community can overcome these barriers and help realise the full potential of affective agents. 


%



\section*{Acknowledgment}
This work has received funding from the DFG's Reinhart Koselleck project No.\ 442218748 (AUDI0NOMOUS).

\ifCLASSOPTIONcaptionsoff
  \newpage
\fi

\section{\refname}
\printbibliography[heading=none]

%
\begin{IEEEbiographynophoto}{Andreas Triantafyllopoulos}
obtained his Diploma in ECE from the University of Patras, Greece, in 2017.
He is now working as a research assistant researcher at the Chair of Embedded Intelligence for Health Care and Wellbeing at the University of Augsburg, where he has been pursuing his doctoral degree since 2018.
His current focus is on deep learning methods for auditory intelligence and affective computing.
He is a Student Member of the IEEE and its Signal Processing Society (SPS).
\end{IEEEbiographynophoto}

\begin{IEEEbiographynophoto}{Bj\"orn W.\ Schuller}
(M'06 -- SM'15 -- F'18) received his diploma, doctoral degree, habilitation, and Adjunct Teaching Professor all in EE/IT from TUM in Munich/Germany. 
He is Full Professor of Artificial Intelligence and the Head of GLAM at Imperial College London/UK, Full Professor and Chair of Embedded Intelligence for Health Care and Wellbeing at the University of Augsburg/Germany, co-founding CEO and current CSO of audEERING.
He is a Fellow of the IEEE and Golden Core Awardee of the IEEE Computer Society, Fellow of the BCS, Fellow of the ISCA, President-Emeritus of the AAAC, and Senior Member of the ACM. 
He (co-)authored 1,200+ publications (45k+ citations, h-index=100+), is Field Chief Editor of Frontiers in Digital Health and was Editor in Chief of the IEEE Transactions on Affective Computing amongst manifold further commitments and service to the community. 
\end{IEEEbiographynophoto}

\begin{IEEEbiographynophoto}{G\"ok\c{c}e \.{I}ymen}
received her B.\,S.\ degree in Industrial Engineering from Middle East Technical University, Turkey in 2019. She is currently studying for a M.\,S.\ degree in Data Science at Koç University, Turkey, where her research focuses on applications of deep learning for audio generation, especially adding affect to speech.
\end{IEEEbiographynophoto}

\begin{IEEEbiographynophoto}{Metin Sezgin}
graduated summa cum laude with Honors from Syracuse University in 1999. He completed his MS in the Artificial Intelligence Laboratory at Massachusetts Institute of Technology in 2001. He received his PhD in 2006 from Massachusetts Institute of Technology. 
Dr.\ Sezgin is currently an Associate Professor in the College of Engineering at Koç University, Istanbul. 
His research interests include intelligent human-computer interfaces, multimodal sensor fusion, and HCI applications of machine learning. 
He has held visiting posts at Harvard University and Yale University. 
His research has been supported by international and national grants including grants from the European Research Council, and Turk Telekom. 
He is a recipient of the Career Award of the Scientific and Technological Research Council of Turkey. 
\end{IEEEbiographynophoto}

\begin{IEEEbiographynophoto}{Xiangheng He}
is a PhD candidate with GLAM -- the Group on Language, Audio, \& Music, Imperial College London, London, UK. She is currently also a research assistant with University of Augsburg, Germany. She received her Master degree from Southeast University, China, in 2020. Her research focuses on affective computing and voice conversion.
\end{IEEEbiographynophoto}

\begin{IEEEbiographynophoto}{Zijiang Yang}
(Student Member, IEEE) received his Master degree in Information Technology from University of York, UK, in 2016. He is currently a research assistant and pursuing his Ph.\,D.\ degree with University of Augsburg, Germany. His research focuses on deep learning, affective computing and speech synthesis.
\end{IEEEbiographynophoto}

\begin{IEEEbiographynophoto}{Panagiotis Tzirakis}
earned his Ph.\,D.\ with the Intelligent Behaviour Understanding
Group (iBUG) at Imperial College London, where he focused on
multimodal emotion recognition efforts. 
He has published in top outlets including Information Fusion, International Journal of
Computer Vision, and several IEEE conference proceedings on
topics including 3D facial motion synthesis, multi-channel speech
enhancement, the detection of Gibbon calls, and emotion
recognition from audio and video.
\end{IEEEbiographynophoto}

\begin{IEEEbiographynophoto}{Shuo Liu}
received his 
M.\,Sc.\ degree in electric engineering and
information technology from Technical University
of Darmstadt (TUD), Darmstadt, Germany, in 2017.
He is currently pursuing his doctoral degree with
the Chair of Embedded Intelligence for Health Care
and Wellbeing at the University of Augsburg. 
His current research interests include deep
learning and machine learning algorithms for speech
and audio processing, affective computing, and health-related applications.
\end{IEEEbiographynophoto}

\begin{IEEEbiographynophoto}{{Silvan Mertes}}
is a PhD candidate at the Chair of Human-Centered Artificial Intelligence at Augsburg University in Germany, where he also received his M.\,Sc.\ degree in Computer Science in 2019. His research focuses on Generative Adversarial Learning for audio and image synthesis. Specifically, he explores how adversarial learning approaches can enhance datasets and explainability in different deep learning tasks.
\end{IEEEbiographynophoto}

\begin{IEEEbiographynophoto}{Elisabeth Andr{\'e}}
is a full professor of Computer Science and Founding Chair of Human-Centered Artificial Intelligence at Augsburg University in Germany, and co-speaker of the Bavarian Research Association ForDigitHealth. 
She has a long track record in multimodal human-machine interaction, embodied conversational agents, social robotics, affective computing and social signal processing. 
Her work has won many awards including the Gottfried Wilhelm Leibniz Prize, the most important research funding award in Germany, and she is a member of the prestigious Academy of Europe, the German Academy of Sciences Leopoldina and the CHI Academy.
In 2013, she was awarded a EurAI fellowship (European Association for Artificial Intelligence). 
Most recently, she was named one of the 10 most influential figures in the history of AI in Germany by National Society for Informatics (GI). 
Since 2019, she is serving as the Editor-in-Chief of IEEE Transactions on Affective Computing.
\end{IEEEbiographynophoto}

\begin{IEEEbiographynophoto}{{Ruibo Fu}}
is an assistant professor in the National Laboratory of Pattern Recognition, Institute of Automation, Chinese Academy of Sciences, Beijing. He obtained his B.\,E.\ from Beijing University of Aeronautics and Astronautics in 2015 and Ph.\,D.\ from the Institute of Automation, Chinese Academy of Sciences in 2020. His research interests lie with speech synthesis and transfer learning. He has published more than 10 papers in international conferences and journals such as ICASSP and INTERSPEECH and has won the best paper award twice in NCMMSC 2017 and 2019. He won the first prize in personalised speech synthesis competition held by the Ministry of Industry and Information Technology twice in 2019 and 2020. He also won the first prize in the ICASSP2021 Multi-Speaker Multi-Style Voice Cloning Challenge (M2VoC) Challenge.
\end{IEEEbiographynophoto}

\begin{IEEEbiographynophoto}{Jianhua Tao} 
received his PhD degree from Tsinghua
University in 2001, and got his Ms from
Nanjing University in 1996. He is currently a Professor
in NLPR, Institute of Automation, Chinese
Academy of Sciences. His current research interests
include speech synthesis and coding methods, human
computer interaction, multimedia information
processing and pattern recognition. He has published
more than eighty papers on major journals and
proceedings including IEEE Trans.\ on ASLP, and
received several awards from the fields' important conferences,
such as Eurospeech, or NCMMSC. He serves as the chair or program
committee member for several major conferences, including ICPR, ACII,
ICMI, ISCSLP, or NCMMSC. He also serves as the steering committee
member for IEEE Transactions on Affective Computing, associate editor for
Journal on Multimodal User Interface and International Journal on Synthetic
Emotions, and Deputy Editor-in-chief for Chinese Journal of Phonetics.
\end{IEEEbiographynophoto}




\end{document}